\documentclass[aps,10pt,pre,twocolumn,superscriptaddress]{revtex4-1}
\usepackage{graphicx}
\usepackage{float}
\usepackage{amsmath}
\usepackage{amsfonts} 
\usepackage{amssymb}
\usepackage{mathtools} 
\usepackage{color}
\usepackage{xr}
\usepackage[dvipsnames]{xcolor}
\usepackage{tabularx}

\usepackage[english]{babel}
\usepackage{silence}
\usepackage[T1]{fontenc}
\usepackage[utf8]{inputenc}
\usepackage{hyperref}
\hypersetup{colorlinks=true,urlcolor=blue,citecolor=blue,linkcolor=blue}
\newcolumntype{Y}{>{\centering\arraybackslash}X}

\usepackage{balance}

\begin{document}
\title{Thermodynamic stability of hard sphere crystals in dimensions 3 through 10}
\date{\today}
\author{Patrick Charbonneau}
\affiliation{Department of Chemistry, Duke University, Durham, North Carolina 27708}
\affiliation{Department of Physics, Duke University, Durham, North Carolina 27708}
\author{Caitlin M. Gish}
\affiliation{Department of Physics, University of South Florida, Tampa, Florida 33620}
\author{Robert S. Hoy}
\affiliation{Department of Physics, University of South Florida, Tampa, Florida 33620}
\author{Peter K. Morse}
\affiliation{Department of Chemistry, Duke University, Durham, North Carolina 27708}

\maketitle

\section{Abstract}
Although much is known about the metastable liquid branch of hard spheres--from low dimension $d$ up to ${d\to\infty}$--its crystal counterpart remains largely unexplored for $d>3$. In particular, it is unclear whether the crystal phase is thermodynamically stable in high dimensions and thus whether a mean-field theory of crystals can ever be exact. In order to determine the stability range of hard sphere crystals, their equation of state  is here estimated from numerical simulations, and fluid-crystal coexistence conditions are determined using a generalized Frenkel-Ladd scheme to compute absolute crystal free energies. The results show that the crystal phase is stable at least up to $d=10$, and the dimensional trends suggest that crystal stability likely persists well beyond that point.

\section{Introduction}
Although the phase behavior of three-dimensional hard spheres was initially debated, for now more than half a century it has been under solid numerical control~\cite{battimelli_computer_2020}. As density increases, the liquid branch reaches the liquid-crystal coexistence point, and then splits into a thermodynamically stable crystal branch and a metastable fluid branch. Further densifying the latter gives rise to glasses and eventually to jammed solids~\cite{charbonneau_glass_2017}. As dimension $d$ increases, these processes are now fairly well understood~\cite{parisi_mean-field_2010,parisi_theory_2020}, thanks to the liquid structure then steadily simplifying~\cite{charbonneau_geometrical_2012, charbonneau_geometrical_2013,charbonneau_hopping_2014,mangeat_quantitative_2016}. In low dimensions, however, not only does the local structure markedly impact the metastable liquid properties, it even facilitates crystal nucleation~\cite{van_meel_geometrical_2009,van_meel_hard-sphere_2009}. Because increasing $d$ generally promotes glass formation at the expense of crystallization \cite{skoge_packing_2006, van_meel_hard-sphere_2009,charbonneau_numerical_2010}, relatively little is known about what happens to the stable crystal branch for $d>3$.
Whether this branch persists in the limit $d\to\infty$, and whether one can obtain any insight into this limit by considering finite-$d$ systems, remain unclear. The present work aims to shed at least some light on these physical questions.

The primary difficulty of pursuing such a program is that each dimension is endowed with its own particular densest packed (and thus thermodynamically preferred) crystal structure. Previous computational studies~\cite{skoge_packing_2006, van_meel_hard-sphere_2009} have shown that the liquid-crystal coexistence pressure of hard spheres increases with dimension, thus suggesting that the crystal becomes steadily less favorable than the liquid as $d$ increases. This analysis, however, was pursued only over a fairly small dimensional range, and further did not take into account the natural dimensional scaling of the properties of dense liquids. It was furthermore done without proper finite-size scaling considerations, a concerning issue in higher $d$, wherein computational constraints on system sizes are particularly acute. Questions thus remain as per the robustness of this proposal. Moreover, low-dimensional crystals of hard spheres all have relatively similar physical properties, whereas higher $d$ crystals can exhibit exotic features, such as non-trivial zero modes~\cite{conway_sphere_1998}. 

Determining equilibrium conditions for phase coexistence generally implies equating temperature $T$, pressure $P$, and chemical potential $\mu$ in all phases present. Temperature being an irrelevant state variable for hard spheres, situating liquid-crystal ($\ell$-$s$) coexistence reduces to finding $P_\mathrm{\ell} = P_\mathrm{s}=P^\mathrm{coex}$ and $\mu_\mathrm{\ell} = \mu_\mathrm{s}=\mu^\mathrm{coex}$. Through numerical simulations this determination can be straightforwardly achieved by thermodynamically integrating the equation of state, given a reference free energy for each phase. For hard spheres, the virial expansion provides the liquid equation of state with high precision over a broad $d$ range~\cite{clisby_ninth_2006, bishop_equation_2005, lue_molecular_2006, zhang_computation_2014}, and the ideal gas offers a convenient reference state. The core computational difficulty is for the crystal phase. Its equation of state has been only phenomenologically described (via numerical simulations in low $d$), and the reference crystal free energy must be obtained from specialized simulation schemes such as that proposed by Frenkel and Ladd~\cite{frenkel_new_1984,frenkel_understanding_2001,khanna_free_2021}. 

In this work, we report the crystal equation of state and the fluid-crystal coexistence conditions for the densest sphere packings in $d=3$-9, which are obtained from the Bravais lattices $D_3$ (face-centered cubic), $D_4$, $D_5$, $E_6$, $E_7$, $E_8$, and $\Lambda_9$, respectively, as well as for the densest packing in $d=10$, which is obtained from the (non-Bravais-lattice) Best packing, $P_{10c}$~\cite{conway_sphere_1998, best_binary_1980, 2dCrystal, bravais}. Added care is given to the consideration of $\Lambda_9$, which is a laminated lattice composed of two interpenetrating $D_9$ sub-lattices with nontrivial zero modes associated with internal translational degrees of freedom. This case is particularly informative about higher-dimensional crystals, because such modes are present in many of the other $\Lambda$ lattices and $P$ binary codes (but not $P_{10c}$), which describe most of densest known sphere packings in $9\le d \le 29$. (The exception is the Coxeter-Todd  $K_{12}$ lattice for $d=12$~\cite{coxeter_extreme_1953}.) 
 Over the accessible dimensional range, we find that  the freezing density $\varphi_f$ remains well below the (avoided) dynamical transition at $\varphi_d$ and that the melting density $\varphi_m$ roughly tracks but also remains below $\varphi_d$. We additionally obtain an upper bound on the low-density crystal stability, $\varphi_s^\mathrm{min}$, in each dimension, and find that $\varphi_s^\mathrm{min}>\varphi_f$.

The plan for the rest of this article is as follows. Section~\ref{sec:crystalDef} defines each of the lattices considered and their embedding in simulations boxes under periodic boundary conditions. Section~\ref{sec:monteCarlo} describes how the liquid and crystal equations of state are obtained. Section~\ref{sec:freeEnergy} details the calculation of reference state free energies, which leads to phase coexistence results being obtained and described in Sec.~\ref{sec:coexConditions}. Section~\ref{sec:conclusion} briefly summarizes the results and describes possible future research directions. 

\section{Generating and Embedding High-$d$ crystals}
\label{sec:crystalDef}
\begin{figure}
\includegraphics[width=0.5\columnwidth]{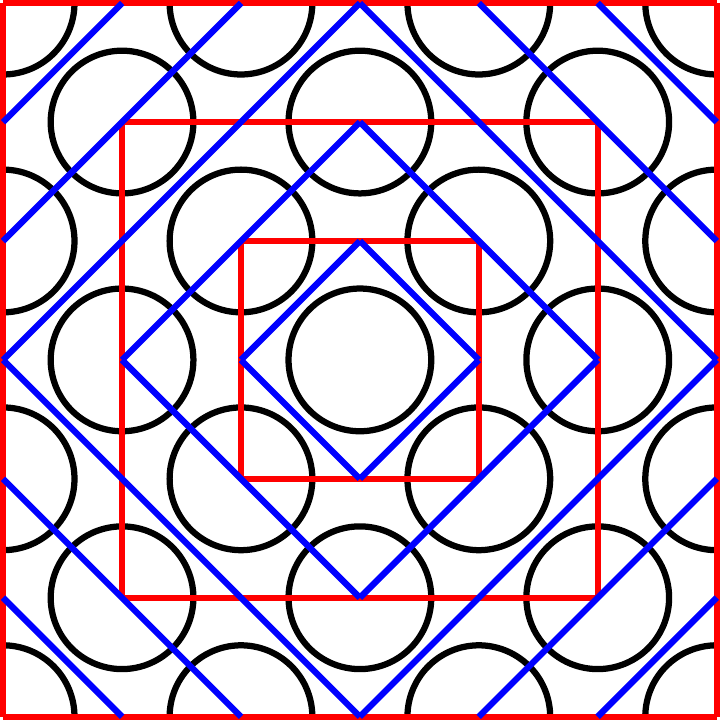}
\caption{Schematic depiction of commensurability between the simulation box the crystal symmetry and for disks on a $D_2$ lattice. Allowed periodic boxes for $Z_2$ (red) boundary conditions and $D_2$ (blue) boundary conditions differ. Periodic boxes are shown for $n = 1$, 2, and 3 in $Z_2$ boundary conditions and $n = 1$, 2, 3, 4, and 5 for $D_2$ boundary conditions.  Clearly, $Z_d$ boundary conditions generically allow fewer system sizes than $D_d$ boundary conditions. Note that while this illustration provides the correct intuition for commensurability in higher dimensions, $d=2$ is a special case because $D_2$ boundary conditions are merely a rotated version of $Z_2$. No such degeneracy exists in higher $d$.}
\label{fig:bcCompare}
\end{figure}

Densest sphere packings in $d=3-9$ are either related to $D$-family (checkerboard) lattices or to the $E_8$ lattice, while the densest sphere packing in $d=10$ is a non-Bravais-lattice packing derived from a binary code~\cite{conway_sphere_1998}. More specifically, we have:
\begin{itemize}
\setlength\itemsep{\fill}
\item the $d$-dimensional $D_d$ (or checkerboard) lattices contain all points $\{x_1, x_2, \dots, x_d\}$ such that $x_i \in \mathbb{Z}$ and ${\sum x_i}$ is an even number;

\item the $E_8$ lattice contains two $D_8$ lattices offset by the eight-dimensional vector $(\frac{1}{2},\ldots,\frac{1}{2})$, such that ${E_8 = D_8 \cup (D_8 + (\frac{1}{2})^8)}$;

\item the $E_7$ lattice is a seven-dimensional subset of $E_8$ consisting of points ${\{x_1, x_2, \dots, x_8\} \in E_8}$ with  ${\sum x_i = 0}$;

\item the $E_6$ lattice is a six-dimensional subset of $E_8$ consisting of points ${\{x_1, x_2, \dots, x_8\} \in E_8}$ with ${\sum x_i = 0}$ and ${x_1 + x_8 = 0}$. Note that the choice of indices $1$ and $8$ is arbitrary, but must be kept consistent;

\item the $D_9^{0+}$  lattice, which is a specific instance from the continuum of $\Lambda_9$ lattices, is analogous to $E_8$, in that it consists of two $D_9$ lattices offset by a vector ${\Xi = \{(\frac{1}{2})^8,\Xi_9\}}$, where $\Xi_9 \in \mathbb{R}$;

\item the Best packing, $P_{10c}$, is defined as the set of all points ${\{x_1, x_2, \dots, x_{10}\} = 2\{a_1, a_2, \dots, a_{10}\} + A^j_{10}}$ where $a_i \in \mathbb{Z}$ and $A^j_{10}$ denotes column $j$ of the 40-column $A_{10}$ matrix~\cite{best_binary_1980}.
\end{itemize}
\begin{widetext}
\begin{equation}
A_{10}=
\begin{bmatrix}
0 & 0 & 1 & 1 & 1 & 1 & 1 & 1 & 0 & 1 & 0 & 1 & 1 & 0 & 0 & 1 & 0 & 1 & 1 & 0 & 1 & 0 & 0 & 0 & 1 & 0 & 1 & 0 & 1 & 1 & 1 & 1 & 0 & 1 & 0 & 0 & 0 & 0 & 0 & 0\\
0 & 1 & 1 & 0 & 0 & 0 & 0 & 0 & 0 & 1 & 1 & 1 & 0 & 0 & 1 & 1 & 0 & 1 & 0 & 0 & 0 & 0 & 1 & 1 & 1 & 0 & 1 & 0 & 1 & 0 & 1 & 0 & 0 & 1 & 0 & 1 & 1 & 1 & 1 & 1\\
0 & 1 & 0 & 0 & 1 & 1 & 1 & 1 & 1 & 1 & 1 & 0 & 0 & 1 & 1 & 0 & 0 & 1 & 0 & 1 & 1 & 1 & 1 & 0 & 0 & 0 & 1 & 0 & 1 & 0 & 0 & 0 & 1 & 1 & 0 & 1 & 0 & 0 & 0 & 0\\
0 & 1 & 0 & 1 & 1 & 0 & 0 & 0 & 0 & 0 & 0 & 0 & 1 & 1 & 0 & 0 & 1 & 1 & 0 & 1 & 1 & 0 & 0 & 0 & 1 & 1 & 1 & 0 & 1 & 0 & 1 & 1 & 1 & 0 & 0 & 1 & 0 & 1 & 1 & 1\\
0 & 0 & 1 & 0 & 1 & 1 & 0 & 0 & 0 & 0 & 1 & 0 & 0 & 1 & 1 & 0 & 0 & 1 & 1 & 0 & 0 & 1 & 0 & 0 & 0 & 1 & 1 & 1 & 0 & 1 & 1 & 1 & 1 & 1 & 0 & 0 & 1 & 0 & 1 & 1\\
0 & 0 & 0 & 1 & 0 & 1 & 1 & 0 & 0 & 0 & 0 & 1 & 0 & 0 & 1 & 1 & 0 & 0 & 1 & 1 & 1 & 0 & 1 & 0 & 0 & 0 & 1 & 1 & 1 & 0 & 1 & 1 & 1 & 1 & 1 & 0 & 0 & 1 & 0 & 1\\
0 & 0 & 0 & 0 & 1 & 0 & 1 & 1 & 0 & 0 & 1 & 0 & 1 & 0 & 0 & 1 & 1 & 0 & 0 & 1 & 0 & 1 & 0 & 1 & 0 & 0 & 0 & 1 & 1 & 1 & 1 & 1 & 1 & 1 & 1 & 1 & 0 & 0 & 1 & 0\\
0 & 0 & 0 & 0 & 0 & 1 & 0 & 1 & 1 & 0 & 1 & 1 & 0 & 1 & 0 & 0 & 1 & 1 & 0 & 0 & 1 & 0 & 1 & 0 & 1 & 0 & 0 & 0 & 1 & 1 & 0 & 1 & 1 & 1 & 1 & 1 & 1 & 0 & 0 & 1\\
0 & 0 & 0 & 0 & 0 & 0 & 1 & 0 & 1 & 1 & 0 & 1 & 1 & 0 & 1 & 0 & 0 & 1 & 1 & 0 & 1 & 1 & 0 & 1 & 0 & 1 & 0 & 0 & 0 & 1 & 1 & 0 & 1 & 1 & 1 & 1 & 1 & 1 & 0 & 0\\
0 & 1 & 1 & 1 & 1 & 1 & 1 & 0 & 1 & 0 & 1 & 1 & 0 & 0 & 1 & 0 & 1 & 1 & 0 & 0 & 0 & 0 & 0 & 1 & 0 & 1 & 0 & 1 & 1 & 1 & 1 & 0 & 1 & 0 & 0 & 0 & 0 & 0 & 0 & 1\\
\end{bmatrix}
\end{equation}
\end{widetext}
The points defined by each of the above are used as sphere center positions to build the crystal. In the units implied by the distances above, spheres of radius $\sigma/2=1/\sqrt{2}$ yield systems at the crystal close packing density, $\varphi_c$. Note that, without loss of generality, we here use $\Xi_9 = \frac{1}{2}$ to build the $d=9$ crystal, but the freedom in selecting $\Xi_9$ translates the existence of an internal zero mode. Additionally, our choice to vary $\Xi_9$ rather than one of the other components of $\Xi$ is arbitrary, and thus the initial choice of the global degree of freedom is itself nine-fold degenerate. 
For these reasons $D_9^{0+}$ requires special consideration in numerical simulations as further discussed in Sec.~\ref{sec:periodicRef}.

Although these crystal definitions may seem straightforward, periodic boundary considerations--which are central to numerical simulations--lead to some geometrical challenges in finding finite-size crystal lattices commensurate with the chosen simulation box. Because only configurations that align with the underlying boundaries are permitted, allowed system sizes, $N$, are sparse, which presents a numerical hurdle in extrapolating results to the thermodynamic limit, $N\rightarrow\infty$. In this context, embedding crystals inside both standard cubic and various non-cubic boundary conditions conveniently shrinks the size gap between commensurate systems. As discussed in Sec.~\ref{sec:crystalFE}, finite-size corrections are indeed largely independent of box shape, provided that shape remains (nearly) isotropic. 

To see how each crystal can be generated within different boundary conditions, consider first standard $Z_d$ symmetric (cubic) simulations boxes. These boxes naturally accommodate $D_3$, $D_4$, $D_5$, $E_8$, and $D_9^{0+}$, which are derived from integer lattices that have themselves $Z_d$ symmetry. Embedding $E_8$ and $D_9^{0+}$ in a $Z_d$ box also relies on embedding a $D_d$ lattice, because these lattices simply fill deep holes in $D_8$ and $D_9$, respectively~\cite{conway_sphere_1998}. In practice, the box is chosen to lie on a $d$-dimensional hypercubic grid with coordinates $[-n,n)$ for $n \in \mathbb{N}$--thus preventing spurious double counting of a sphere and its periodic image--and each grid point is populated with a sphere if it obeys the even sum rule described above. The resulting commensurability condition is illustrated in Fig.~\ref{fig:bcCompare}, which corresponds to $D_d$ lattices of sizes $N = \frac{1}{2}(2n)^d$. For $E_8$ and $D_9^{0+}$, a second lattice is obtained by duplicating and shifting every particle, thus doubling the particle count to $N = (2n)^d$.

Starting from the cubic embedding of any of the above lattices, it is possible to devise an embedding into $D_d$ boundary conditions, and, as a special case, to embed $E_8$ into $E_8$ boundary conditions. This construction is here achieved by choosing the $D_d$ (or $E_8$) boundaries to also have limits $[-n,n)$ that correspond with the blue periodic boxes in Fig.~\ref{fig:bcCompare}. Conway's decoding  algorithms~\cite{conway_fast_1982,charbonneau_dimensional_2021} can then be used to find the spheres that lie outside of the boundary and to map their periodic images back into the box. (Duplicate particles are then straightforwardly removed.) This construction results in  $D_d$ lattices in $D_d$ boundary conditions with $N = n^d$, and $E_8$ lattices in $E_8$ with $N = n^8$. As with the $Z_d$ boundary conditions, $E_8$ and $D_9^{0+}$ can be embedded in $D_d$ boundary conditions by simply embedding $D_8$ or $D_9$, respectively, and then inserting a shifted copy of the lattice. This process creates $E_8$ and $D_9^{0+}$ lattices with  $N = 2n^d$. Note that although, in principle, any $n$ is allowed for these constructions, adequately simulating systems larger than a few tens of thousands of particles falls beyond the reach of commonly available computational resources.

Embedding $E_6$ and $E_7$ in periodic boxes is slightly less straightforward, but can nevertheless be achieved via generating matrices. An embedding of $E_7$ crystals in a $Z_7$ cell follows from the generating matrix (transpose) given by Ref.~\onlinecite{conway_sphere_1998}, using all points
\begin{equation}
\mathbf{r} = 
\begin{bmatrix}
1 & 0 & 0 & 0 & \frac{1}{2} & 0 & 0 \\
0 & 1 &  0 & 0 & \frac{1}{2} & \frac{1}{2} & 0 \\
0 & 0 &  1 & 0 & \frac{1}{2} & \frac{1}{2} & \frac{1}{2} \\
0 & 0 &  0 & 1 & 0 & \frac{1}{2} & \frac{1}{2} \\
0 & 0 & 0 & 0 & \frac{1}{2} & 0 & \frac{1}{2} \\
0 & 0 & 0 & 0 & 0 & \frac{1}{2} & 0 \\
0 & 0 & 0 & 0 & 0 & 0 & \frac{1}{2} 
\end{bmatrix}
\cdot
\begin{bmatrix}
x_1\\
x_2\\
x_3\\
x_4\\
x_5\\
x_6\\
x_7\\
\end{bmatrix}.
\label{eq:E7g}
\end{equation}
with $x_i \in \mathbb{Z}$ lying inside the $\mathbb{Z}_7$ cube with integer side lengths $n\geq 2$. This embedding thus produces systems with $N = 8n^7$.

Although $E_6$ cannot be embedded in $Z_6$ boundary conditions, it can be embedded in nearly-cubic orthorhombic cells using the \emph{simple-root} generating matrix (transpose)~\cite{conway_sphere_1998}. A first such embedding consists of all points
\begin{equation}
\mathbf{r} = \begin{bmatrix}
1 & 0 & 0 & 0 & 0 & -\frac{1}{2} \\
-1 & 1 & 0 & 0 & 0 & -\frac{1}{2} \\
0 & -1 & 1 & 0 & 0 & -\frac{1}{2} \\
0 & 0 & -1 & 1 & 1 & -\frac{1}{2} \\
0 & 0 & 0 & -1 & 1 & -\frac{1}{2} \\
0 & 0 & 0 & 0 & 0 & \frac{\sqrt{3}}{2} 
\end{bmatrix}
\cdot
\begin{bmatrix}
x_1\\
x_2\\
x_3\\
x_4\\
x_5\\
x_6\\
\end{bmatrix}.
\label{eq:E6g}
\end{equation}
with $x_i \in \mathbb{Z}$ lying inside of an orthorhombic cell with side lengths ${\{1,1,1,\frac{\sqrt{3}}{2},1,1\}} n$, where $n\geq 4$ and must be even. This embedding produces systems with $N = n^6/2$. A second such embedding has side lengths ${\{3,3,3,\sqrt{3},\sqrt{3},\sqrt{3}\}}n$ or ${\{3,3,3,2\sqrt{3},2\sqrt{3},2\sqrt{3}\}}$ each containing 24 atoms~\cite{van_meel_hard-sphere_2009}. Both types are used in this work.

$P_{10c}$ is based on a binary code and is thus naturally embedded in $Z_{10}$ boundary conditions~\cite{conway_sphere_1998} via the hypercubic box with coordinates $[0,2n)$ for $n \in \mathbb{N}$. Each unit cell contains 40 particles, yielding $N=40n^{10}$. In order to ensure that each particle has a set of unique neighbors, it is necessary to choose $n \ge 3$, which creates systems too large for us to consider in this present work. However, because the integer lattice itself can be embedded in both $D_{10}$ and $D_{10}^+$ boundary conditions, so too can $P_{10c}$, from simply populating the integer lattice. The integer lattice $Z_d$ can be embedded in the $D_d$ lattice with coordinates $[-n,n)$ for $n \in \mathbb{N}$, creating system sizes $N=2n^{d}$. Here, it is only necessary to choose $n \ge 2$ for each particle to have a unique set of neighbors. Similarly, $Z_d$ can be embedded in the  $D_d^{+}$ (for $d$ even) or $D_d^{0+}$ (for $d$ odd) lattice with coordinates $[-n,n)$ for $n \in \mathbb{N}$, creating system sizes $N=n^{d}$. Thus, $P_{10c}$ can be embedded in $D_{10}$ with $N=80n^{10}$ and in $D_{10}^+$ with $N=40n^{10}$.

\section{Liquid and Crystal Equations of State}
This section describes the computational and analytical approaches used to determine the fluid and crystal equations of state.
\label{sec:monteCarlo}
\begin{figure*}
\includegraphics[width=\linewidth]{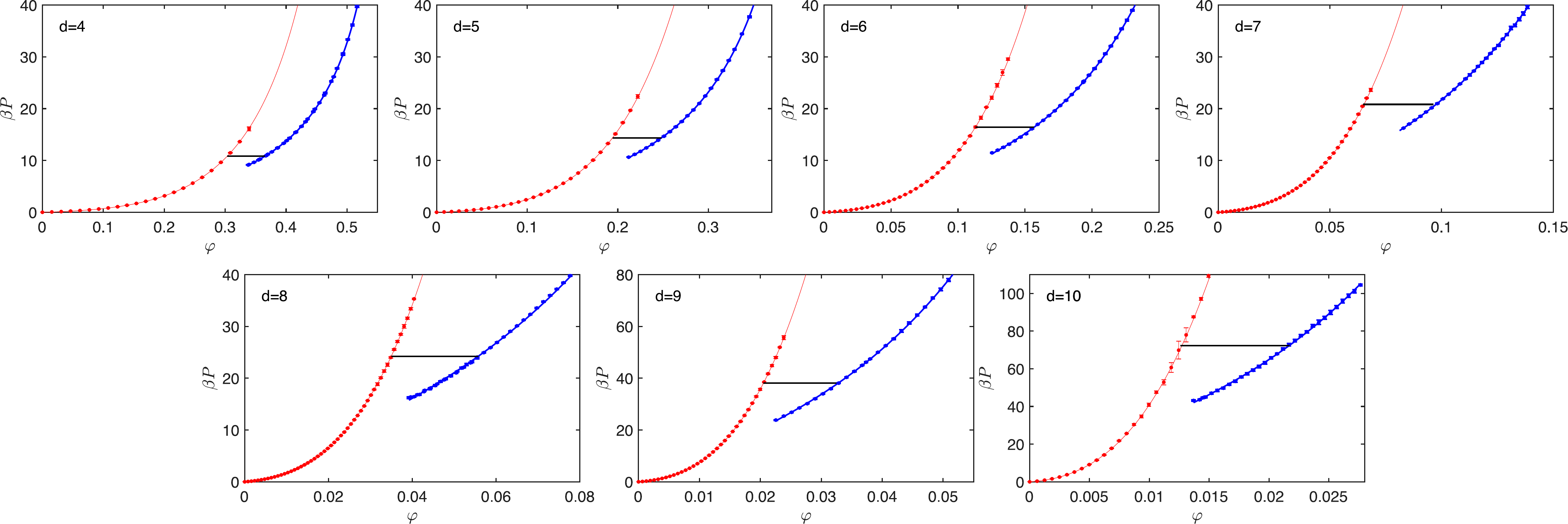}
\caption{Liquid (red) and crystal (blue) equations of state and coexistence conditions (black) in $d=4$-$10$. Pressures from numerical simulations (points) of the liquid are obtained for $N = 2048$, 3888, 3000, 3000, 1000, 9000, and 6000 in $d=4$-$10$ respectively (where $d=8$-$10$ are embedded in $E_8$, $D_9$, and $D_{10}$ respectively) and are in quantitative agreement with the [4,5] Pad\'e approximant (line) in $d=4$-$9$ and the [5,6] Pad\'e approximant in $d=10$ of the virial expansion (Eq.~\eqref{eqn:liquidEOS})~\cite{clisby_ninth_2006, bishop_equation_2005, lue_molecular_2006, zhang_computation_2014}. Pressures from numerical simulations (points) of the crystal are obtained for 2048, 3888, 5184, 17496, 6561, 39366, and 81920 in $d=4$-$10$ respectively. Fits to the empirical form given in Eq.~\eqref{eqn:crystalEOS} (solid blue line) quantitatively recapitulate the simulation results. Coexistence conditions are such that $\mu_\ell = \mu_s$ and $P_\ell = P_s$, as described in Sec.~\ref{sec:coexConditions}. Error bars denote 95\% confidence intervals. Pressure is reported in units that implicitly set the particle diameter to unity, i.e., with $\sigma=1$.}
\label{fig:eos}
\end{figure*}

\subsection{Monte Carlo Simulations}
Equilibrium configurations are sampled using a standard Metropolis Monte Carlo (MC) scheme, which defines the unit of time $t$ as one MC cycle. For liquids, we define the structural relaxation time, $\tau_\alpha$, as the characteristic decay time of the standard overlap parameter for the chosen MC dynamics~\cite{frenkel_understanding_2001}, and for crystals, we define $\tau_\alpha$ as the characteristic time needed for the mean squared displacement to reach its plateau~\cite{charbonneau_three_2021}. In both cases, systems are deemed equilibrated for simulations run for $t>10\tau_\alpha$, and 10,000 independent configurations are generated for each density. For the $D_9^{0+}$ crystal the overall equilibration parameters remain the same but additional MC sampling moves are used to accelerate the sampling of its zero modes. More specifically, the $\Xi_d$ degrees of freedom are sampled by using MC moves which displace one $D_9$ sub-lattice with respect to the other. At the point of maximum degeneracy, ${\Xi = \{(\frac{1}{2})^9\}}$, nine such pairs of lattices can be created, one for each $\Xi_d$ chosen as the global degree of freedom. Thus, for $N$ single particle moves and $1$ center of mass displacement (as motivated in Sec.~\ref{sec:periodicRef}), $9$ relative $D_9$ subset moves are used, on average, for each MC cycle. In order to preserve the symmetry of the governing Markov chain, each MC move is given equal weight, and thus we sample each MC move with frequency $1/(N+10)$. These moves are essential for efficiently sampling the $D_9^{0+}$ crystal, as they yield a speedup of at least $10^4$ over standard MC (as estimated from the pressure equilibration, or rather the lack thereof).

\subsection{Analytical Forms}
\label{sec:EOS}
Pressure $P$ is extracted from the radial distribution function, $g(r)$. The virial theorem gives the reduced pressure (also called the compressibility)
\begin{equation}
p(\rho) = \frac{\beta P}{\rho} = 1 + 2^{d-1}\varphi g(\sigma^+)
\end{equation}
at packing fraction ${\varphi = \rho \sigma^d V_d}$, where $V$ is the box volume, $V_d$ is the $d$-dimensional volume of a sphere of unit diameter, $\beta$ is the inverse temperature, and ${\rho = \frac{N}{V}\sigma^d}$ is the number density of spheres of diameter $\sigma$. The value of the pair correlation function at contact, $g(\sigma^+)$, is obtained by extrapolating a quadratic fit of nearby $g(r)$ results. Note that the reduced pressure is often denoted $Z$, but we here follow the convention of Ref.~\onlinecite{parisi_theory_2020} to avoid notational collision with the contact number. Following standard conventions, all distances are reported for a unit particle diameter, i.e., $\sigma=1$.

For the liquid in $d\leq 9$, the equation of state is well approximated by the $[4,5]$ Pad\'e approximant of the virial series
\begin{equation}
p = \frac{1 + \sum_{i=1}^4 b_i\rho^i}{\sum_{i=1}^5 \bar{b}_i\rho^i}.
\label{eqn:liquidEOS}
\end{equation}
with coefficients $b_i$ and $\bar{b}_i$~\cite{bishop_equation_2005, lue_molecular_2006} obtained from the first 10 virial coefficients computed by Clisby and McCoy \cite{clisby_ninth_2006}. More terms are needed for $d=10$, hence we use the $[5,6]$ Pad\'e approximant, obtained by resumming higher order virial coefficients~\cite{zhang_computation_2014}. As can be seen in Fig.~\ref{fig:eos}, these forms fall well within the 95\% confidence intervals of the numerical results, at least up to the fluid-crystal coexistence regime. For $d=3$, in order to obtain an even higher accuracy, we follow the high precision work of Pieprzyk et al.~on $\approx 10^6$ particles~\cite{pieprzyk_thermodynamic_2019} and use a simple polynomial form that is accurate up to $\varphi=0.534$.

For the crystal, various equations of state have been proposed~\cite{hall_another_1972, speedy_pressure_1998, tarazona_density_2000} with careful numerical studies lending greatest credence to that of Speedy in $d=3$~\cite{speedy_pressure_1998}. Here, the Speedy form with the high-accuracy coefficients of Pieperzyk et al.~is used for $d=3$~\cite{pieprzyk_thermodynamic_2019}. This form, however, is ill-conditioned in higher dimensions. For our purpose, given that the reference point (see Sec.~\ref{sec:crystalFE}) is by construction close to the melting density, a simple second-order polynomial correction to the free volume scaling suffices,
\begin{equation}
p_s = \frac{1}{1-\big(\frac{\varphi}{\varphi_c}\big)^{1/d}} + a_0 + a_1 \bigg(\frac{\varphi}{\varphi_c}\bigg) + a_2 \bigg(\frac{\varphi}{\varphi_c}\bigg)^2,
\label{eqn:crystalEOS}
\end{equation}
where coefficients $a_0$, $a_1$, and $a_2$ are determined by fitting the numerical reduced pressure results. For a given density, the pressure of the liquid is higher than that of the crystal. When the crystal density is lowered below its lowest (meta)stable density, the system pressure then rises once a stable liquid manages to nucleate. The lowest stable crystal density $\varphi_s^\mathrm{min}$ is here estimated as the lowest density at which pressure does not rise over times comparable to $\tau_\alpha$. 

\section{Free Energy Determination}
\label{sec:freeEnergy}


Given the equation of state, free energy differences can be obtained by thermodynamic integration. In order to obtain absolute values, a reference point of known free energy must be available. The efficient selection of such a point depends on the nature of the phase considered. This section describes the various schemes used in this work.

\subsection{Fluid free energy}
In the fluid, the (Helmholtz) free energy can be computed using the ideal gas as reference state. The thermodynamic integration can then be written as
\begin{equation}
\frac{\beta F_\mathrm{\ell}(\rho)}{N} = \frac{\beta F_\mathrm{id}(\rho)}{N} + \int_0^\rho \frac{p - 1}{\rho'}d\rho'
\label{eqn:solidTI}
\end{equation}
with the ideal gas free energy given by
\begin{equation}
\frac{\beta F_\mathrm{id}}{N} = \mathrm{ln}(\rho\Lambda^d) - 1 + \frac{\ln(2 \pi N)}{2N}.
\end{equation}
Without loss of generality, we set the de Broglie wavelength to unity, $\Lambda = 1$ .

\subsection{Crystal free energy}
\label{sec:crystalFE}
In the crystal phase, a similar thermodynamic integration is possible, 
\begin{equation}
\frac{\beta F_\mathrm{s}(\rho)}{N} = \frac{\beta F_\mathrm{s}(\rho_0)}{N} + \int_{\rho_0}^\rho \frac{p}{\rho'} d\rho',
\end{equation}
albeit using a system of known free energy at number density, $\rho_0$, as reference. (For numerical efficacy, this density is chosen near an estimate of the melting density.) The free energy of such reference crystal is obtained by Frenkel-Ladd integration from a model crystal which is exactly solvable~\cite{frenkel_new_1984}.

Given a reference crystal with energy $U_0$ and using a coupling parameter $\lambda\in[0,\infty)$, we can write the energy of an alchemical system as
%
%
\begin{equation}
U(\lambda) =  U_\mathrm{HS} + \lambda U_0
\label{eqn:totalEnergy}
\end{equation}
where $U_\mathrm{HS}$ is the hard sphere potential, which is recovered for $\lambda=0$. The reference free energy can then be generically written as
\begin{equation}
\frac{\beta F_s(\rho_0)}{N} = -\frac{\beta}{N}\int^\lambda_0 \langle U \rangle_\lambda d\lambda \\ + \frac{\beta F_\mathrm{corr}(\rho_0)}{N},  
\label{eqn:einsteinThermoInt}
\end{equation}
%
where the subscript $\lambda$ denotes a thermal average taken at constant $\lambda$, and a correction term is added as per Ref.~\cite{polson_finite-size_2000}. In the limit $\lambda\rightarrow\infty$, $U_0$ dominates and the system free energy can be approximated from its contribution alone. In practice, setting a large $\lambda_\mathrm{max}$, allows the integral to be separated as
\begin{equation}
\int^\lambda_0 \langle U\rangle_\lambda d\lambda = \\ F(\lambda>\lambda_\mathrm{max}) +\int^{\lambda_\mathrm{\rm max}}_0 \langle U_0\rangle_\lambda d\lambda,
\label{eqn:TISplit}
\end{equation}
%
where $\lambda_\mathrm{max}$ can be estimated analytically for simple reference systems~\cite[Eq.~10.3.31]{frenkel_understanding_2001}. Note that because $\lambda_\mathrm{max}\gg1$, it is numerically convenient to change integration coordinates as
\begin{equation}
\int^{\lambda_\mathrm{max}}_0 \langle U_0\rangle_\lambda d\lambda= \lim_{\lambda_\mathrm{min}\rightarrow0^+} \int^{\ln(\lambda_\mathrm{max})}_{\ln(\lambda_\mathrm{min})} \lambda \langle U_0\rangle_\lambda d(\ln\lambda).
\end{equation}
The remaining problem is to find a reference crystal whose free energy can be explicitly calculated. The standard solution is to consider an Einstein crystal in which each particle $i$ at position $\mathbf{r}_{i}$ is harmonically tethered to its perfect crystal site, $\mathbf{r}_{0,i}$ 
\begin{equation}
U_0 =  \sum^N_{i=1}(\mathbf{r}_i - \mathbf{r}_{0,i})^2.
\label{eqn:einsteinEnergy}
\end{equation}
For a non-interacting Einstein crystal the free energy ${F_\mathrm{Ein} = F(\lambda>\lambda_\mathrm{max})}$ is given by
\begin{equation}
\frac{\beta F_\mathrm{Ein}(\rho_0)}{N} = \frac{d}{2} \ln\bigg({\frac{\pi}{\lambda_{\rm max}}}\bigg) + \frac{d}{2N} \ln\bigg(\frac{N\lambda_{\rm max}}{\pi}\bigg),
\label{eqn:einsteinFE}
\end{equation}
with the correction term~\cite{polson_finite-size_2000}
\begin{equation}
\frac{\beta F_\mathrm{corr}(\rho_0)}{N} = - \frac{\ln{\rho_0}}{N} + \frac{d-1}{2}\frac{\ln N}{N}.
\end{equation}
Note that the center of mass contribution (first term) is here treated separately, because it must be kept fixed in numerical simulations. The integrand, ${\langle U\rangle_\lambda = \langle r^2 \rangle_\lambda}$, is the mean squared displacement of a system equilibrated with an energy given by Eq.~\eqref{eqn:einsteinEnergy}, which indeed does not permit system-wide translations. In Section~\ref{sec:periodicRef}, we consider an alternative reference crystal given by a periodic potential whose center of mass is not fixed, and is therefore better suited for the study of $\Lambda_9$. 

%
\begin{figure}
\includegraphics[width=0.75\columnwidth]{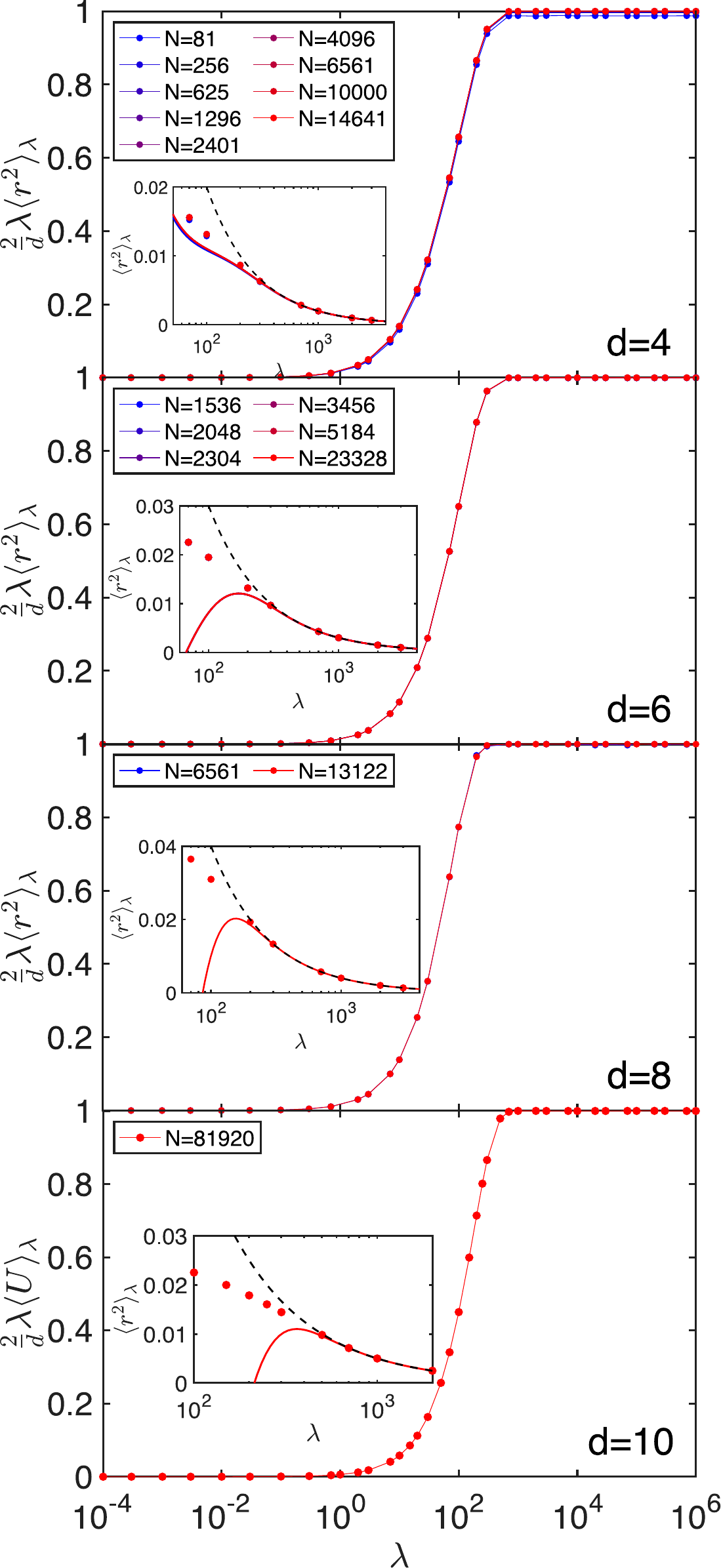}
\caption{Evolution of the Einstein crystal integrand $\langle U \rangle$ (Eq.~\eqref{eqn:einsteinThermoInt}) with tethering constant $\lambda$ in $d = 4$, $6$, $8$, and $10$. Multiplying the integrand by $2/d$ accounts for the trivial dimensional scaling. Note that finite-size corrections for each integrand scale as $1/N$, and are thus indistinguishable on this scale. For the integration, we choose $\lambda_{\rm min} = 10^{-4}$, below which the function is essentially 0, and $\lambda_{\rm max} \approx 1400$ (determined for each crystal using the analytical approach of Ref.~\cite[Eq.~10.3.31]{frenkel_understanding_2001}), such that deviations from an Einstein crystal result in a relative error ${1-\langle r^2\rangle_\lambda/\langle r^2 \rangle_{\mathrm{Ein},\lambda} \ll 1/N}$. These corrections are thus numerically negligible. Insets show the mean squared displacement in the Einstein crystal simulations along with the analytic result for the interacting (red line) and non-interacting (black dashed line) Einstein crystal in the large $\lambda$ limit~\cite[Eq.~10.3.30]{frenkel_understanding_2001}).}
\label{fig:RMSEinstein}
\end{figure}

\begin{figure}
\includegraphics[width=\columnwidth]{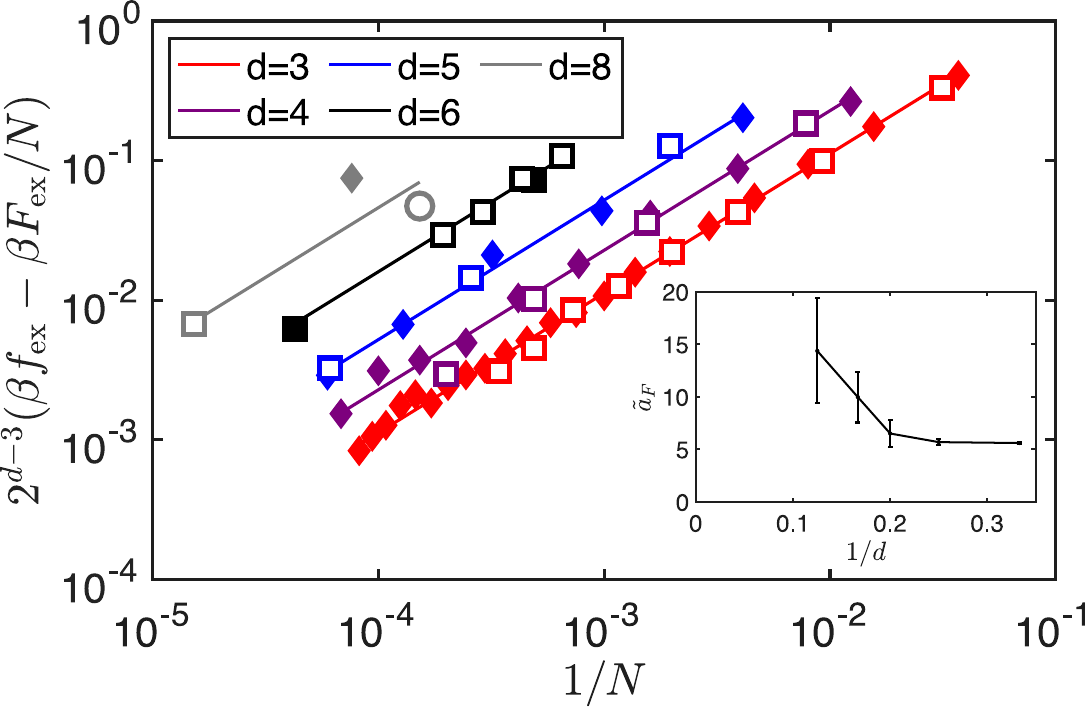}
\caption{Reference crystal free energies for $Z_d$ (squares), $D_d$ (diamonds) and $E_8$ (circles) embeddings. For $d=6$, full and empty squares denote orthorhombic boxes of the first and second types of $E_6$ embedding, respectively. Because results for a given $d\leq 6$ are essentially colinear, we conclude that (sufficiently isotropic) boundary conditions have little effect on the first-order correction to the intensivity of the free energy in this size regime. All results are thus used to fit Eq.~\eqref{eqn:finSizeF} (lines), and for the ensuing thermodynamic extrapolation. The sole outlier is $E_8$, for which only the smallest box in each embedding type is computationally accessible and is thus more prone to preasymptotic corrections. The larger relative error on the extrapolated thermodynamic free energy reflects this fact. Error bars for 95\% confidence intervals are smaller than the symbols and are thus neglected for visual clarity. Curves are offset by a multiplicative factor of $2^{d-3}$ for visual clarity. \textbf{(inset)} Dimensional evolution of $\tilde{a}_F$ from Eq.~\eqref{eqn:finSizeF}. }
\label{fig:coexistenceLogLog}
\end{figure}

Numerical integration of the integrand (using a cubic spline smoothing function) chooses $\lambda_\mathrm{min}$ such that contributions for $\lambda < \lambda_\mathrm{min}$ are negligible, and $\lambda_\mathrm{max}$ such that the remainder $\lambda > \lambda_\mathrm{max}$ approaches the result for a non-interacting crystal.
In practice (see Fig.~\ref{fig:RMSEinstein}), setting $\lambda_\mathrm{min} \approx 10^{-4}$ and $\lambda_\mathrm{max} \approx 1400$ suffices. As $d$ increases, the first-order correction used to estimate $\lambda_\mathrm{max}$ has a smaller convergence radius, but above $\lambda_\mathrm{max}$ the integrand plateaus, hence that portion of the integral can be approximated as $\beta F_\mathrm{Ein}/N$, as expected. Note also that the integrand varies smoothly and monotonically, which validates the choice and implementation of the integration scheme. 

As Polson et al.~have argued~\cite{polson_finite-size_2000}, the reference free energy scales asymptotically as $1/N$.  We thus fit the numerical results to
\begin{equation}
\frac{\beta F(\varphi_0)}{N} = \beta f(\varphi_0) - \frac{\tilde{a}_f}{N}
\label{eqn:finSizeF}
\end{equation}
to extrapolate the thermodynamic limit of the reference free energy per particle, $f(\varphi_0)$ (Fig.~\ref{fig:coexistenceLogLog} and Table~\ref{tab:refFE}). In $d=3$-$6$ and $d=8$ the number of points used for the fit is $\ge 3$, hence a statistical error can be reported for $\beta f$. For $d=7$, $9$, and $10$ only one system size is studied, hence a different error estimate must be obtained. Here, we note that the error on $\beta f$ is the quadrature sum of the error of $\beta F/N$ and $\tilde{a}_f/N$. Thus, if $\tilde{a}_f$ is the primary source of error and it is bound from above, then a maximum error estimate on $\beta f$ can be obtained by assuming that the error in $\tilde{a}_f$ is itself equal to the maximum bound. Although the scaling constant $\tilde{a}_f$ increases with $d$, it is divided by $N$ in Eq.~\eqref{eqn:finSizeF}, and the smallest crystal sizes considered in these high dimensions are still rather large. The extrapolation error for $\beta f$ is thus expected to remain small even in $d=10$. To provide a quantitative estimate, we guess that $\tilde{a}_f \approx 10$ in $d=7$, $\tilde{a}_f \approx 100$ in $d=9$, and $\tilde{a}_f \approx 300$ in $d=10$.

With Eq.~\eqref{eqn:finSizeF} and the crystal equation of state from Eq.~\eqref{eqn:crystalEOS}, thermodynamic integration in Eq~\eqref{eqn:solidTI} is then used to determine the crystal free energy for conditions near coexistence.

\bgroup
\begin{table}
\caption{Reference crystal excess free energies, $\beta f_\textrm{ex} = \beta f - \beta f_\textrm{id}$, extrapolated in the thermodynamic limit from Eq.~\eqref{eqn:finSizeF} with errors denoting 95\% confidence intervals in $d=3$-$6$ and $8$, and estimated assuming $\tilde{a}_f \leq 10$ for $d=7$, $\tilde{a}_f \leq 100$ for $d=9$, and $\tilde{a}_f \leq 300$ for $d=10$.}
\def\arraystretch{1.2}
\begin{tabular}{c | c | c }
\hline
\hline
$d$ & $\varphi_0$ & $\beta f_\textrm{ex}(\varphi_0)$ \\
\hline
3 & 0.5450 & 5.9188(3)\\
4 & 0.34 & 6.2869(4)\\
5 & 0.21 & 7.3840(6)\\
6 & 0.14 & 8.7828(8)\\
7 & 0.087 & 9.806(1)\\
8 & 0.048 & 9.818(15)\\
9 & 0.0322& 11.92(3)\\
10 & 0.0229 & 15.569(4) \\
\hline
\hline
\end{tabular}

\label{tab:refFE}
\end{table}
\egroup

\subsection{Periodic Potential Crystal Reference}
\label{sec:periodicRef}

As noted in Sec.~\ref{sec:crystalDef}, the global internal zero modes along the various $\Xi_9$ axis of $\Lambda_9$ crystals require special consideration. An Einstein crystal reference is then inappropriate because particle displacements cannot be bounded.
To surmount this issue, we here draw inspiration from the simulation of the crystal phase of parallel cubes. In order to account for the rich collection of zero modes in this model, Groh et al.~\cite{groh_closer_2001} proposed using a periodic potential crystal that matches the symmetry of the crystal phase of interest as reference.

For example,
for all $D_d$ lattices we define the external potential $U_0$, 
\begin{equation}
U_0 = \frac{w^2}{8\pi^2}\sum_i^N\bigg(1-\prod_\alpha^d \cos\big[\frac{2\pi}{w} (\mathbf{r}_i \cdot \hat{x}_\alpha) \big]\bigg)
\label{eqn:einPeriodic}
\end{equation}
where $\mathbf{r}_i$ is the position of particle $i$, $\hat{x}_\alpha$ is the unit vector in the $\alpha$ direction, and $w=2\sigma(\varphi_c/\varphi)^{1/d}$ is the lattice spacing. 

By contrast to Einstein crystals (Eq.~\eqref{eqn:einsteinEnergy}), the crystal center of mass should not be kept fixed but should instead thoroughly sample the system volume. Dedicated center of mass displacements are thus incorporated at a frequency of $1/N$ relative to individual particle MC moves (Ref.~\cite[Sec.~3.3]{frenkel_understanding_2001}). The finite-size scaling of the free energy is also duly modified, 
\begin{equation}
\frac{\beta F_s(\rho_0)}{N} = -\frac{d}{2} \ln\bigg({\frac{\pi}{\lambda_\mathrm{max}}}\bigg) - \frac{\beta}{N}\int^{\lambda_\mathrm{max}}_0 \langle U_0\rangle_\lambda d\lambda ,
\label{eqn:FEPeriodic}
\end{equation}
where the first term is the free energy of the reference periodic potential crystal, obtained by Taylor expanding around the minimum. By construction, it is equal to the first term in Eq.~\eqref{eqn:einsteinFE}. Note that the second term in that equation, which accounts for the fixed center of mass, is not here present because the center of mass is now unconstrained.

\begin{figure}
\includegraphics[width=0.95\linewidth]{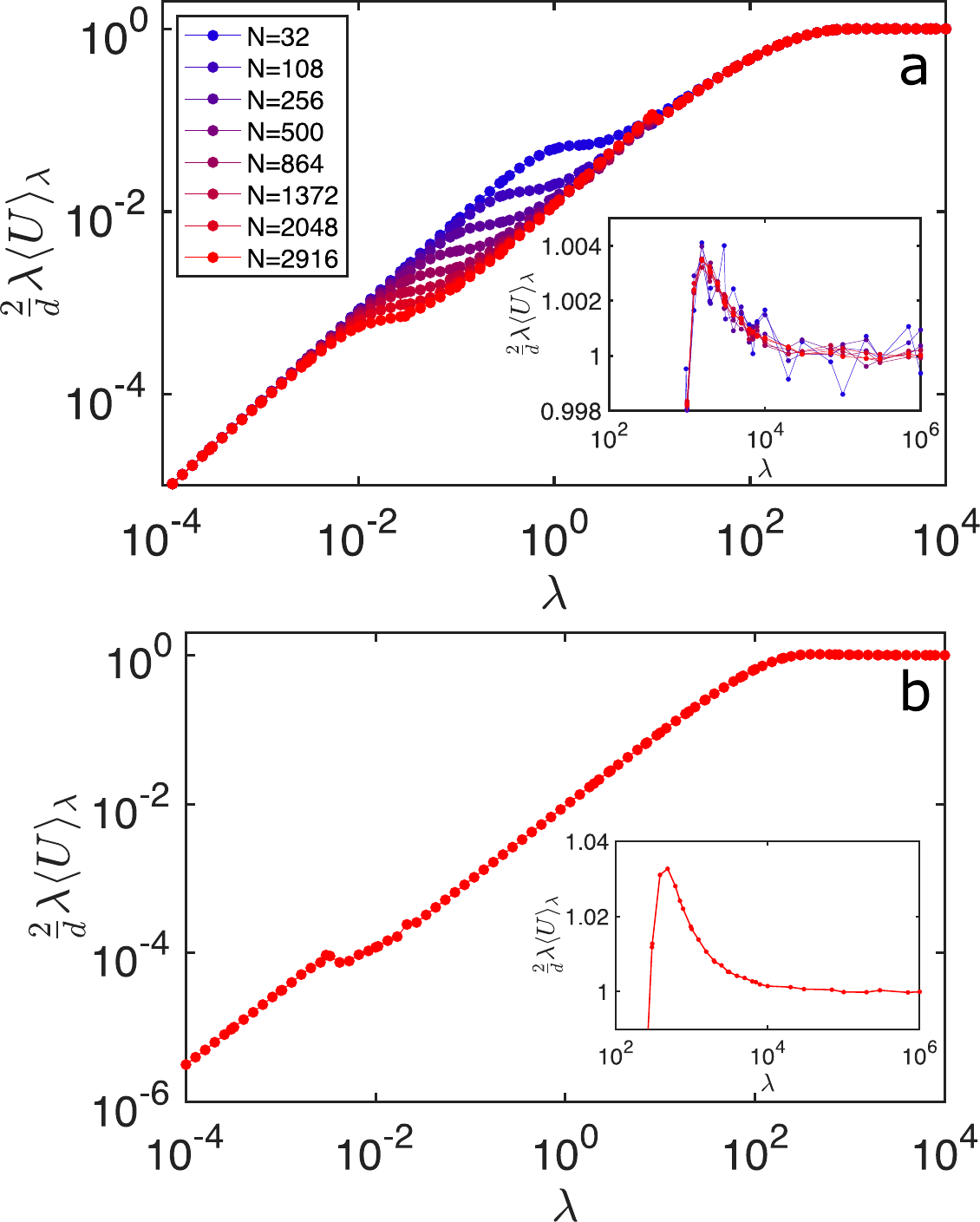}
\caption{The integrand of Eq.~\eqref{eqn:FEPeriodic} in \textbf{a)} $d=3$ and \textbf{b)} $d=9$ using a periodic reference field on a log-log axis. Umbrella sampling for $\lambda < 1$ ensures that the position of the center of mass samples uniformly the box volume. This enhanced sampling scheme is particularly important near the lower shoulder of the curve, at which point the center of mass gradually untethers from the minima of the external field. (Inset) Close-up on the large $\lambda$ regime, showing the nonmonotonicity of the integrand, and validating the choice $\lambda_\mathrm{max}=10^6$ for this system.}
\label{fig:periodicIntegrand}
\end{figure}

Validating this approach against the Einstein crystal reference free energies in $d=3$-5 reveals that a direct application of the periodic potential yields a region of the integrand of Eq.~\eqref{eqn:FEPeriodic} that is particularly challenging to sample. The crossover from unimpeded to limited center of mass displacement is indeed associated with rapidly changing capability to thermally sample energy barriers. An umbrella sampling scheme~\cite{torrie_nonphysical_1977,frenkel_understanding_2001} is thus used to compensate for this difficulty. Given a non-negative weighting function $\theta(\mathbf{r}^N)$, the ensuing Markov chain distribution 
\begin{equation}
\pi(\mathbf{r}) = \frac{\theta(\mathbf{r})\exp[-\beta U_0(\mathbf{r})]}{\int \theta(\mathbf{r'})\exp[-\beta U_0(\mathbf{r'})]d^d\mathbf{r'}} 
\label{eqn:umbrellaMarkov}
\end{equation}
results in weighted averages of standard thermodynamic quantities.  Specifically,
\begin{equation}
\langle U_0 \rangle_\lambda = \frac{\langle U_0/\theta \rangle_{\lambda,\pi}}{\langle 1/\theta \rangle_{\lambda,\pi}} ,
\label{eqn:umbrellaAverage}
\end{equation}
where $\langle\dots\rangle_{\lambda\theta}$ denotes an average taken at constant $\lambda$ in the $\pi$-weighted ensemble. From Eq.~\eqref{eqn:umbrellaMarkov}, a particularly convenient choice of the weighting function is one that provides an equal probability of being in all states by exactly canceling the Boltzmann weight of the field,
\begin{equation}
\theta(\mathbf{r}) = e^{\beta U_0(\mathbf{r})} . 
\label{eqn:umbrellaWeight}
\end{equation}
This choice results in the position of the center of mass sampling uniformly the whole box volume. Note that although this scheme is appropriate (and efficient) for $\lambda < 1$, the weighting function becomes nearly singular for $\lambda \gg 1$ (see Appendix~\ref{sec:singleParticle}). Note also that the crossover from $\lambda \lesssim1$ to $\lambda \gg 1$ creates a slight overshoot of the integrand of Eq.~\eqref{eqn:FEPeriodic} (Fig.~\ref{fig:periodicIntegrand} inset). In practice, robust numerical results are obtained by combining the standard periodic potential for $\lambda \ge 1$ with the umbrella sampling scheme for $\lambda < 1$.

For the $D_9^{0+}$ lattice, an additional term is introduced in Eq.~\eqref{eqn:einPeriodic} to account for the second sub-lattice,
\begin{multline}
U_0 = \frac{w^2}{8\pi^2}\sum_i^N\bigg(1-\prod_\alpha^d \cos\big[\frac{2\pi}{w} (\mathbf{r}_i \cdot \hat{x}_\alpha) \big] \bigg) \cdot \\
\bigg(1-\prod_\alpha^d\cos\big[\frac{2\pi}{w} ([\mathbf{r}_i - \Xi] \cdot \hat{x}_\alpha) \big] \bigg),
\label{eqn:einPeriodicD9}
\end{multline}
where ${\Xi = \{\Xi_d\} = \{(\frac{1}{2})^9\}}$ is the chosen offset.

\bgroup
\begin{table*}
\caption{Liquid-crystal coexistence parameters for various $d$ reported with 95\% confidence intervals, along with reference values for $\varphi_d$ ($d=3$-8~\cite{charbonneau_hopping_2014}, $d=9$~\cite{charbonneau_dimensional_2021}, and $d=10$~\cite{charbonneau_glass_2011}) and $\varphi_c$~\cite{conway_sphere_1998}. Errors for $\varphi_s^\mathrm{min}$ are set by the simulation point resolution.
}
\label{tab:coexInf}
\def\arraystretch{1.2}
\begin{tabularx}{0.7\linewidth}{ c | Y | Y | Y | Y | Y | Y | Y}
\hline
\hline
$d$ & $P^\mathrm{coex}$ & $\mu^\mathrm{coex}$ & $\varphi_f$ & $\varphi_m$ & $\varphi_s^\mathrm{min}$ & $\varphi_d$ & $\varphi_c$  \\
\hline
3 & 11.578(3) & 16.082(4)& 0.4919(3) & 0.5434(4)& 0.5142(16) & 0.5770(5) &0.7405  \\
4 & 10.807(3) & 15.133(3)& 0.3031(3) & 0.3653(3)& 0.324(16) & 0.4036(2) &06169 \\
5 & 14.363(3) & 17.604(3)& 0.1942(1) & 0.2484(3) & 0.206(6) & 0.2683(1) &0.4653 \\
6 & 16.400(4) & 17.697(3)& 0.1129(1)& 0.1567(2) & 0.125(5) & 0.1723(1) &0.3729 \\
7 & 20.43(1) & 18.26(1)&0.0648(1)&0.0963(2)& 0.080(2) & 0.1076(1) &0.2953 \\
8 & 24.23(9) & 17.82(4)&0.0350(5)& 0.0558(2)& 0.0392(11) & 0.06585(5) &0.2537 \\
9 & 38.4(1) & 20.2(1)&0.0206(8)&0.0329(6) & 0.0225(12) & 0.0391(6) & 0.1458 \\
10 & 71.9(1) & 24.6(1) & 0.0126(3) & 0.0216(3) & 0.0137(5) & 0.0226(1) & 0.0996\\
\hline
\hline
\end{tabularx}
\end{table*}
\egroup

\section{Coexistence conditions}
\label{sec:coexConditions}
Given the thermodynamic reference free energy  along with integrals of the equations of state in Eqs.~\eqref{eqn:liquidEOS} and~\eqref{eqn:crystalEOS}, the chemical potential can be obtained as
\begin{equation}
\beta \mu(\rho) = \beta f(\rho) + p(\rho).
\label{eqn:muDef}
\end{equation} 
Coexistence conditions $(P^\mathrm{coex},\mu^\mathrm{coex})$ can then be determined from the crossing point of parameteric plots of $\mu_\ell$ vs $P_\ell$ and $\mu_s$ vs $P_s$. Freezing and melting densities can further be extracted from the liquid and solid equations of state, respectively.

The coexistence results in Table~\ref{tab:coexInf} lie within the range of previously reported values obtained using a variety of different techniques (Table~\ref{tab:compareValues}) for $d<6$. However, they differ significantly from previous results reported by one of us~\cite{van_meel_hard-sphere_2009}, which did not consider finite-size corrections and were numerically quite crude. In $d \ge 6$, our results markedly  differ from previous reports. Because these  estimates did not directly probe the thermodynamics of coexistence, but opted for estimates that have limited first-principle support, a clear physical explanation for the discrepancy is not immediate. A possible explanation is that these approaches might have (fortuitously) worked well in $3 \leq d <6$, for which the lattice family remains unchanged, but that the change in lattice family for $d\geq 6$ was less forgiving. Note that other robust numerical methods for calculating coexistence exist and could be used to test the claims of this work, including phase switch Monte Carlo~\cite{wilding_freezing_2000, wilding_new_2002}, but these are not considered here.
\begin{figure}
\includegraphics[width=0.95\linewidth]{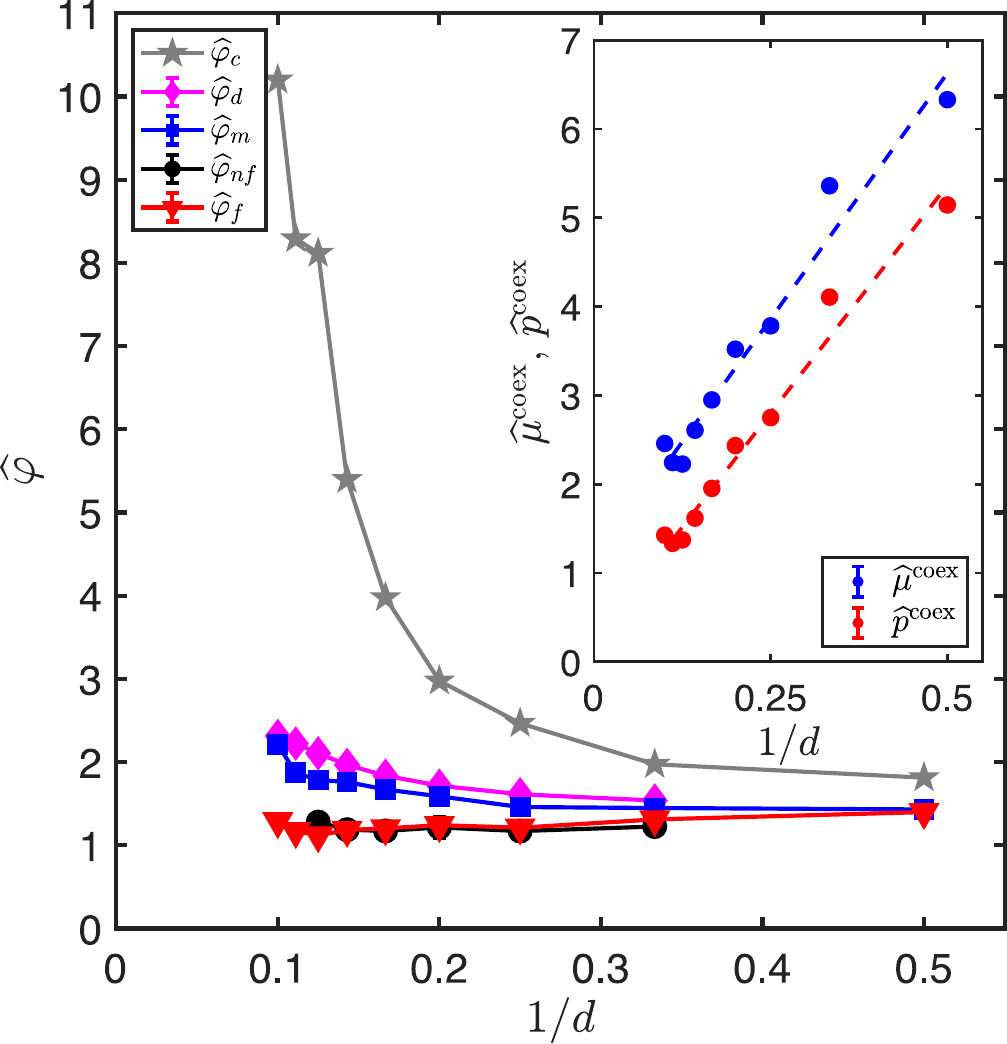}
\caption{Asymptotically scaled freezing $\widehat{\varphi}_f$ and melting $\widehat{\varphi}_m$. While $\widehat{\varphi}_f$ appears to tend to a constant equal to the onset of non-Fickian diffusion $\widehat{\varphi}_{nf}$, $\widehat{\varphi}_m$ steadily increases with $d$, seemingly tracking the (avoided) dynamical glass transition $\widehat{\varphi}_d$ (from Refs.~\onlinecite{charbonneau_glass_2011, charbonneau_dimensional_2021}). Data for $d=2$ are obtained by  combining the liquid-hexatic coexistence density reported in Ref.~\onlinecite{engel_hard-disk_2013} with the liquid equation of state given by the $[4,5]$ Pad\'e approximant of the virial expansion from Ref.~\onlinecite{clisby_ninth_2006}. \textbf{(Inset)} Dimensional evolution of the similarly scaled coexistence pressure $\widehat{p}=\beta P/(\rho_\ell d)$ and chemical potential $\widehat{\mu}$. Dashed lines are guides to the eye.}
\label{fig:coexistencePressureVsD}
\end{figure}

In order to compare densities and pressures across dimensions, we consider and correct for their asymptotic $d\rightarrow\infty$ scaling, using results for the dense liquid state~\cite{parisi_mean-field_2010}. We thus consider rescaled reduced pressures $\widehat{p} = p/d$, rescaled chemical potentials ${\widehat{\mu} = \beta \mu/d}$, and rescaled densities $\widehat{\varphi} = 2^d\varphi/d$. In this form, the melting and freezing densities suggest an interesting physical picture (Fig. \ref{fig:coexistencePressureVsD}a). Even though the lattice close packing density grows rapidly with $d$, the freezing point is nearly constant, $\widehat{\varphi}_f \approx 1.3$. Interestingly, this observation is consistent with the recent phenomenological observation that $\varphi_f$ and the onset of non-Fickian diffusion $\varphi_{nf}$ (nearly) coincide in $d=3$~\cite{ruiz-franco_coincidence_2019}. It also stands in stark contrast with the predictions of MFCT~\cite{wang_mean-field_2005},
\begin{equation}
\widehat{\varphi}_f^\mathrm{MFCT} = \frac{\widehat{\varphi}_c (1+2d)}{2+\widehat{\varphi}_c\frac{2d^2}{V_d}} \xrightarrow{d\to\infty} \frac{V_d}{d} \to 0.
\end{equation}
While the freezing point thus lies well below the (avoided) dynamical transition $\varphi_d$, the melting point $\varphi_m$ approaches, yet remains below, $\varphi_d$ in all $d$ considered. Taken together with the coexistence pressure and chemical potential results, these observations suggest that hard sphere crystallization--albeit rare--is not significantly impeded by the slowdown of the fluid dynamics for $d \le 10$. Using the crude estimate $\widehat{\varphi}_f \approx 1.3$, and the high dimensional equation of state $\widehat{p}=\frac{\widehat{\varphi}}{2}$ we can also extrapolate the $d\rightarrow\infty$ fluid-crystal coexistence conditions as $\widehat{p}^\mathrm{coex} \approx 0.65$. Applying thermodynamic integration in this limit then yields $\widehat{\mu} = 2\widehat{p}$, and thus $\widehat{\mu}^\mathrm{coex} \approx 1.3$. This prediction, however, deviates from the extrapolated lines in Fig.~\ref{fig:coexistencePressureVsD}, implying that, as $d$ increases, either the scalings of the pressure and of the chemical potential become nonlinear or $\widehat{\varphi}_f$ decreases. Without guidance from a proper theory of mean-field crystallization to account for the evolving crystal symmetry with $d$, further speculation remains rather tentative.

This scaling form nevertheless suggests a rough description of the relative stability of crystals across the dimensions considered. For $d \le 9$, odd-$d$ crystals have coexistence pressures and chemical potentials which lie above the trend line, while those of their even-$d$ counterparts lie below it. The latter are therefore relatively more thermodynamically stable than the former, with $\widehat{\mu}^\mathrm{coex}$ for $E_8$ lying notably below the overall trend. This feature likely reflects $E_8$ being in a sense the densest sphere lattice for $d>1$, given its near saturation of the Rogers bound~\cite{rogers_packing_1958} and its actual saturation of the more strict Cohn-Elkies bound~\cite{cohn_new_2003, viazovska_sphere_2017, cohn_conceptual_2017}. 

The coexistence pressure and chemical potential of the non-Bravais-lattice packing $P_{10c}$ fall far above the trend line of the lattices, and perhaps give a more generic case of what should be expected in higher dimensions, where the difference between the Cohn-Elkies bound and the densest known packing increases markedly. However, testing this hypothesis through the simulation of the next several densest crystals ($P_{11a}$, $K_{12}$, and $P_{13a}$ in $d=11$-$13$~\cite{conway_sphere_1998}) would require considerably larger computational resources and is thus left for future work.

Another form of crystal (meta)stability is the resistance to melting below the coexistence pressure. In order for a homogeneous crystal to melt, a  nucleation site must typically form. The free energy barrier that controls this activated process is, however, expected to vanish at a spinodal-like point, below which the crystal is truly unstable. Our simulations provide an upper bound for this last quantity, $\varphi_s^\mathrm{min}$ (Table \ref{tab:coexInf}). We note that although the gap between the freezing density and $\varphi_s^\mathrm{min}$ shrinks with dimension, the two quantities are distinct, and $\varphi_f < \varphi_s^\mathrm{min}$ even in $d=10$. These observations nevertheless suggest that the coexistence estimate used in Ref.~\onlinecite{skoge_packing_2006}, which relies on this instability in $d=4$-$6$, would likely fall far off the mark were it applied to higher $d$ crystals.

\bgroup
\begin{table}
\caption{Previously reported coexistence conditions in order of publication for each dimension. Estrada et al.~\cite{estrada_fluidsolid_2011} obtained an estimate using polynomial fitting (PF) and another using universal relations (UR), denoted accordingly. Values reported without error bars are from publications in which none were provided. Results from the current work are given in bold. 
}
\label{tab:compareValues}
\resizebox{\columnwidth}{!}{
\def\arraystretch{1.2}
\begin{tabular}{c | c | c | c | c | c }
\hline
\hline
$d$ & $P^\mathrm{coex}$ &  $\mu^\mathrm{coex}$ & $\varphi_f$ & $\varphi_m$ & Ref.\\
\hline
3 & 11.70(18) & -- & 0.494(2) & 0.545(2) & \cite{hoover_melting_1968}\\
& -- & -- & 0.487 & -- &\cite{michels_dynamical_1984}\\
& 11.564 & 17.071 & 0.494 & 0.545 & \cite{frenkel_understanding_2001} \\
& 11.55(11) & -- & 0.491(1) & 0.543(1) &\cite{speedy_pressure_1997}\\
& 11.49(9) & -- & 0.489(2) & 0.540(2) & \cite{wilding_freezing_2000}\\
& -- & -- & 0.494 & -- &\cite{alder_phase_1957, bryant_how_2002, wang_mean-field_2005}\\
& 11.54(4) & -- & -- & -- & \cite{vega_revisiting_2007}\\
& 11.202  &--& 0.488(5) & 0.537 & \cite{estrada_fluidsolid_2011} PF\\
& 11.668 &--& 0.492 & 0.542 & \cite{estrada_fluidsolid_2011} UR\\
& 11.5712(10) & 16.0758(20) & 0.49176(5) & 0.54329(5) & \cite{pieprzyk_thermodynamic_2019}\\
& \textbf{11.578(3)} & \textbf{16.082(4)}& \textbf{0.4919(3)} & \textbf{0.5434(4)}&\\
\hline
4 & -- & -- & 0.308 & -- &\cite{michels_dynamical_1984, wang_mean-field_2005}\\
& -- & -- & 0.32(1) & 0.39(1) & \cite{skoge_packing_2006} \\
& 9.15 & 13.7 & 0.288 & 0.337 & \cite{van_meel_hard-sphere_2009} \\
& 11.008 & -- & 0.304(1) & 0.368 & \cite{estrada_fluidsolid_2011} PF\\
& 11.469 & -- & 0.308 & 0.374 & \cite{estrada_fluidsolid_2011} UR\\
& \textbf{10.807(3)} & \textbf{15.133(3)} & \textbf{0.3031(3)} & \textbf{0.3653(3)}&\\
\hline
5 & -- & -- & 0.194 & -- &\cite{michels_dynamical_1984}\\
& -- & -- & 0.169 & -- &\cite{wang_mean-field_2005}\\
& -- & -- & 0.20(1) & 0.25(1) & \cite{skoge_packing_2006} \\
& 10.2 & 14.6 & 0.174 & 0.206 & \cite{van_meel_hard-sphere_2009} \\
& 13.433 & -- & 0.190(1) & 0.242 & \cite{estrada_fluidsolid_2011} PF\\
& 13.184 & -- & 0.189 & 0.240 & \cite{estrada_fluidsolid_2011} UR\\
& \textbf{14.363(3)} & \textbf{17.604(3)} & \textbf{0.1942(1)} & \textbf{0.2484(3)} &\\
\hline
6 & -- & -- & 0.084 & -- &\cite{wang_mean-field_2005}\\
& 13.3 & 16.0 & 0.105 & 0.138 & \cite{van_meel_hard-sphere_2009} \\
& 16.668  & -- & 0.114(2) & 0.146 & \cite{estrada_fluidsolid_2011} PF\\
& 17.0318 & -- & 0.114 & 0.147 & \cite{estrada_fluidsolid_2011} UR\\
& \textbf{16.400(4)} & \textbf{17.697(3)} & \textbf{0.1129(1)} & \textbf{0.1567(2)} &\\
\hline
7 & -- & -- & 0.039 & -- &\cite{wang_mean-field_2005}\\
& 22.597  & -- &0.0702(2) & 0.086 & \cite{estrada_fluidsolid_2011} PF\\
& 22.1569 & -- & 0.0696 & 0.085 & \cite{estrada_fluidsolid_2011} UR\\
& \textbf{20.43(1)} & \textbf{18.26(1)} & \textbf{0.0648(1)} & \textbf{0.0963(2)} &\\
\hline
8 & -- & -- & 0.017 & -- &\cite{wang_mean-field_2005}\\
& -- & -- & 0.0427 & -- & \cite{estrada_fluidsolid_2011} UR\\
& \textbf{24.23(9)} & \textbf{17.82(4)} & \textbf{0.0350(5)} & \textbf{0.0558(2)} &\\
\hline
9 & \textbf{38.4(1)} & \textbf{20.2(1)} & \textbf{0.0206(8)} & \textbf{0.0329(6)} &\\
\hline
10 & \textbf{71.9(1)} & \textbf{24.6(1)} & \textbf{0.0126(3)} & \textbf{0.0216(3)} &\\
\hline
\hline
\end{tabular}
}
\end{table}
\egroup

\section{Conclusion}
\label{sec:conclusion}

From this analysis, it is clear that the crystal phase is thermodynamically stable at high pressures in dimensions $d=3$-$10$. It also appears that $\widehat{p}^\mathrm{coex}$ and $\widehat{\mu}^\mathrm{coex}$ tend smoothly towards finite values in the limit $d\rightarrow\infty$. Given our scheme for the $\Lambda_9$ lattice and relatively weak dimensional dependence of $\tilde{a}_f$, generalizing our study to the non-root lattices that dominate in $d>8$~\cite{conway_sphere_1998} should be conceivable. The minimal system sizes needed for these studies, however, are too computationally prohibitive for the moment. 

With coexistence conditions firmly established in ${d=3}$-10, questions about the dimensional evolution of nucleation and melting nevertheless persist. Because $\widehat{\varphi}_f$ appears to be dimensionally invariant, it remains to be shown what factor actually controls the height of the crystallization barrier in high-dimensional systems. Computing the solid-liquid interfacial free energy~\cite{richard_crystallization_2018, richard_crystallization_2018-1, Bultmann_computation_2020} would further enlighten this trend. This effort is also left for future work.

Finally, our methodological improvements for the periodic potential crystal reference should also find applications in a variety of more common two- and three-dimensional systems. For instance, it could be used to revisit the phase behavior of parallel cubes as well as for exploring that of a number of crystals of polyhedra~\cite{damasceno_predictive_2012} and superballs \cite{batten_phase_2010} that exhibit comparable zero modes.

\begin{acknowledgments}
We thank 	Yi Hu, Irem Altan, and Francesco Zamponi for fruitful discussions. Additionally, we thank Henry Cohn for suggesting to expand this work to $d=10$. This work was supported by grants from the Simons Foundation (\#454937, Patrick Charbonneau) and from the National Science Foundation under Grant DMR-2026271 (Robert Hoy).
The computations were carried out on the Duke Compute Cluster (DCC), for which the authors thank Tom Milledge’s assistance. Data relevant to this work have been archived and can be accessed at the Duke Digital Repository~\cite{data}.

\textbf{Author contributions:} PC, RSH, and PKM designed the research and wrote the manuscript; CMG and PKM performed simulations; PC, CMG, RSH, and PKM analyzed data. Author list is alphabetical.
\end{acknowledgments}

\appendix

\section{Free energy of a single particle in the periodic reference crystal field}
\label{sec:singleParticle}

A minimal model of the periodic potential crystal reference calculation is a single particle evolving in the field given by Eq.~\eqref{eqn:einPeriodic} (see Fig.~\ref{fig:periodicIntegrand}). Because the problem can then be solved analytically, it provides a robust benchmark for our numerical implementation and its optimization.

\begin{figure}
\includegraphics[width=0.95\linewidth]{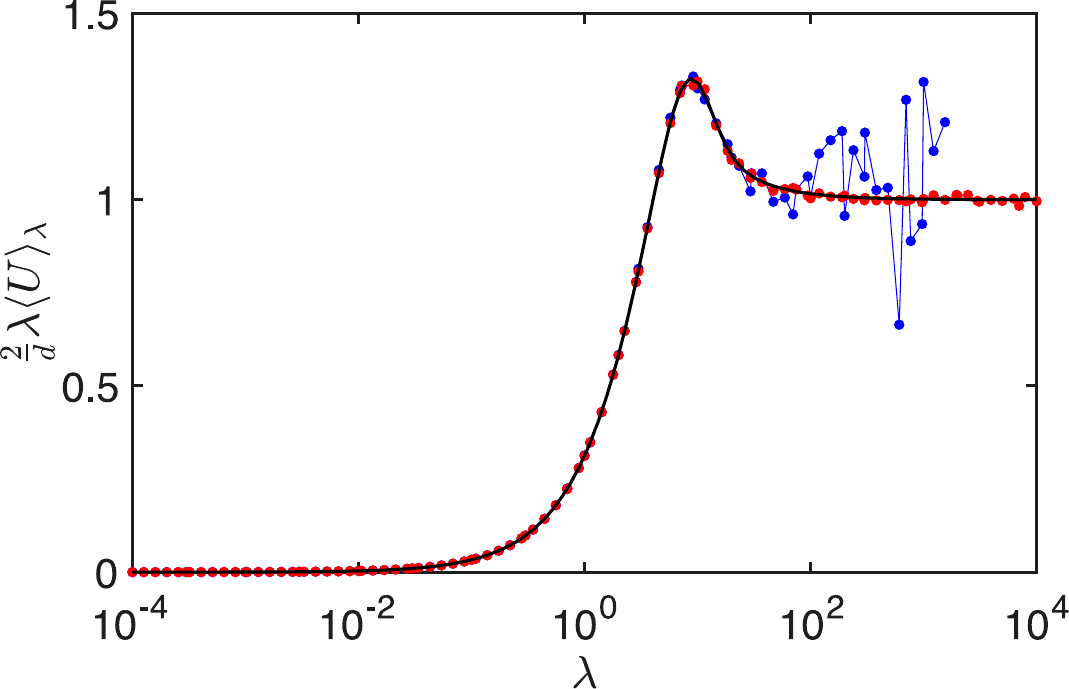}
\caption{Integrand of Eq.~\eqref{eqn:FEPeriodic} for a single particle in the periodic $d=3$ reference crystal from direct integration (black line), direct simulation of the periodic potential (red points), and simulation of the umbrella potential of Eq.~\eqref{eqn:umbrellaAverage} (blue points). The different results are indistinguishable below $\lambda \approx 10$, but for $\lambda \gtrsim 10$ results from the umbrella potential deviate from the correct solution due to poor sampling of the bottom of the potential well, wherein the particle is then largely confined.}
\label{fig:periodicIntegrandSingleParticle}
\end{figure}

The integrand of Eq.~\eqref{eqn:FEPeriodic} can then be written explicitly as 
\begin{equation}
\langle U(\mathbf{r},\lambda')\rangle_{\lambda'} = \frac{\int U(\mathbf{r},\lambda') \exp[-\beta U(\mathbf{r},\lambda')]d^d\mathbf{r}}{\int \exp[-\beta U(\mathbf{r},\lambda')] d^d\mathbf{r}},
\end{equation}
and thus the system free energy is $\beta F = -\beta \int_0^\lambda \langle U(\mathbf{r},\lambda')\rangle_{\lambda'}$. Although the integrand lacks a closed form expression for $d>1$, it can be evaluated numerically with very high accuracy for any $\lambda$ and $d$. The results for $d=3$ serve as an analytical reference in Fig.~\ref{fig:periodicIntegrandSingleParticle}. Equivalently, the free energy can be computed directly from the standard statistical mechanics expression, $\beta F = -\ln Z$, for the partition function $Z$, and hence
\begin{equation}
\beta F = -\ln \int \exp[-\beta U(\mathbf{r},\lambda)] d^d \mathbf{r}.
\end{equation}
This expression also does not have a closed form, but can be calculated numerically with high accuracy. 
 
These quantities can be used to benchmark the standard Monte Carlo sampling of the periodic potential as well as by the umbrella sampling scheme described in Sec.~\ref{sec:periodicRef}. For $\lambda \gg 1$ standard Monte Carlo sampling leaves the particle trapped at the bottom of one of the wells, but as $\lambda$ decreases the particle regularly explores barriers and crosses over into neighboring wells. Given sufficient sampling, the intermediate $\lambda$ regime is recovered (see Fig.~\ref{fig:periodicIntegrandSingleParticle}), and thus the various features of the direct integration are recapitulated. 

This outcome contrasts with Monte Carlo simulations that use the umbrella sampling approach of Eq.~\eqref{eqn:umbrellaWeight}. In this case, the particle samples all points in the box with equal probability, even though the actual contribution to the partition function of each point is proportional to its Boltzmann weight, $\exp[-\lambda \beta U]$. In the single particle case, both simple Monte Carlo and umbrella sampling work well for $\lambda \lesssim 10$. However, for $N>1$ and $\lambda \sim 1/N$ standard Monte Carlo sampling fails because it then becomes rare for  simulations to produce particles which are near the top of energy barriers, whose contributions remain important for accurately calculating the free energy.

For $\lambda \gg 1$, the Boltzmann weight $\exp[-\lambda \beta U]$ concentrates, and the only significant contributions to $Z$ come from a small collection of (degenerate) points in phase space, i.e., the bottom of each well. Both the numerator and the denominator of Eq.~\eqref{eqn:umbrellaAverage} diverge in this limit, making the evaluation numerically unstable. Physically, this instability corresponds to $\langle U \rangle$ being vastly undersampled compared with other $\lambda$ regimes, thus leading to marked deviation from the exact result (Fig.~\ref{fig:periodicIntegrandSingleParticle}).

This analysis motivates the high $\lambda$ cutoff used in the umbrella sampling for $N>1$. Because the Boltzmann weight $\exp[-\lambda \beta \sum_i U_i]$ acts on the total energy, the same cutoff of $\lambda=1$ can be used. However, as $N$ grows, it is important to sum these contributions carefully, because $\theta$ then also grows.

\bibliography{liquidCrystalCoexistence,footnotes}

\begin{thebibliography}{58}%
\makeatletter
\providecommand \@ifxundefined [1]{%
 \@ifx{#1\undefined}
}%
\providecommand \@ifnum [1]{%
 \ifnum #1\expandafter \@firstoftwo
 \else \expandafter \@secondoftwo
 \fi
}%
\providecommand \@ifx [1]{%
 \ifx #1\expandafter \@firstoftwo
 \else \expandafter \@secondoftwo
 \fi
}%
\providecommand \natexlab [1]{#1}%
\providecommand \enquote  [1]{``#1''}%
\providecommand \bibnamefont  [1]{#1}%
\providecommand \bibfnamefont [1]{#1}%
\providecommand \citenamefont [1]{#1}%
\providecommand \href@noop [0]{\@secondoftwo}%
\providecommand \href [0]{\begingroup \@sanitize@url \@href}%
\providecommand \@href[1]{\@@startlink{#1}\@@href}%
\providecommand \@@href[1]{\endgroup#1\@@endlink}%
\providecommand \@sanitize@url [0]{\catcode `\\12\catcode `\$12\catcode
  `\&12\catcode `\#12\catcode `\^12\catcode `\_12\catcode `\%12\relax}%
\providecommand \@@startlink[1]{}%
\providecommand \@@endlink[0]{}%
\providecommand \url  [0]{\begingroup\@sanitize@url \@url }%
\providecommand \@url [1]{\endgroup\@href {#1}{\urlprefix }}%
\providecommand \urlprefix  [0]{URL }%
\providecommand \Eprint [0]{\href }%
\providecommand \doibase [0]{http://dx.doi.org/}%
\providecommand \selectlanguage [0]{\@gobble}%
\providecommand \bibinfo  [0]{\@secondoftwo}%
\providecommand \bibfield  [0]{\@secondoftwo}%
\providecommand \translation [1]{[#1]}%
\providecommand \BibitemOpen [0]{}%
\providecommand \bibitemStop [0]{}%
\providecommand \bibitemNoStop [0]{.\EOS\space}%
\providecommand \EOS [0]{\spacefactor3000\relax}%
\providecommand \BibitemShut  [1]{\csname bibitem#1\endcsname}%
\let\auto@bib@innerbib\@empty
\bibitem [{\citenamefont {Battimelli}\ \emph {et~al.}(2020)\citenamefont
  {Battimelli}, \citenamefont {Ciccotti},\ and\ \citenamefont
  {Greco}}]{battimelli_computer_2020}%
  \BibitemOpen
  \bibfield  {author} {\bibinfo {author} {\bibfnamefont {G.}~\bibnamefont
  {Battimelli}}, \bibinfo {author} {\bibfnamefont {G.}~\bibnamefont
  {Ciccotti}}, \ and\ \bibinfo {author} {\bibfnamefont {P.}~\bibnamefont
  {Greco}},\ }\href {\doibase 10.1007/978-3-030-39399-1} {\emph {\bibinfo
  {title} {Computer {Meets} {Theoretical} {Physics}: {The} {New} {Frontier} of
  {Molecular} {Simulation}}}},\ The {Frontiers} {Collection}\ (\bibinfo
  {publisher} {Springer International Publishing},\ \bibinfo {year}
  {2020})\BibitemShut {NoStop}%
\bibitem [{\citenamefont {Charbonneau}\ \emph {et~al.}(2017)\citenamefont
  {Charbonneau}, \citenamefont {Kurchan}, \citenamefont {Parisi}, \citenamefont
  {Urbani},\ and\ \citenamefont {Zamponi}}]{charbonneau_glass_2017}%
  \BibitemOpen
  \bibfield  {author} {\bibinfo {author} {\bibfnamefont {P.}~\bibnamefont
  {Charbonneau}}, \bibinfo {author} {\bibfnamefont {J.}~\bibnamefont
  {Kurchan}}, \bibinfo {author} {\bibfnamefont {G.}~\bibnamefont {Parisi}},
  \bibinfo {author} {\bibfnamefont {P.}~\bibnamefont {Urbani}}, \ and\ \bibinfo
  {author} {\bibfnamefont {F.}~\bibnamefont {Zamponi}},\ }\href {\doibase
  10.1146/annurev-conmatphys-031016-025334} {\bibfield  {journal} {\bibinfo
  {journal} {Annu. Rev. Condens. Matter Phys.}\ }\textbf {\bibinfo {volume}
  {8}},\ \bibinfo {pages} {265} (\bibinfo {year} {2017})}\BibitemShut {NoStop}%
\bibitem [{\citenamefont {Parisi}\ and\ \citenamefont
  {Zamponi}(2010)}]{parisi_mean-field_2010}%
  \BibitemOpen
  \bibfield  {author} {\bibinfo {author} {\bibfnamefont {G.}~\bibnamefont
  {Parisi}}\ and\ \bibinfo {author} {\bibfnamefont {F.}~\bibnamefont
  {Zamponi}},\ }\href {\doibase 10.1103/RevModPhys.82.789} {\bibfield
  {journal} {\bibinfo  {journal} {Rev. Mod. Phys.}\ }\textbf {\bibinfo {volume}
  {82}},\ \bibinfo {pages} {789} (\bibinfo {year} {2010})}\BibitemShut
  {NoStop}%
\bibitem [{\citenamefont {Parisi}\ \emph {et~al.}(2020)\citenamefont {Parisi},
  \citenamefont {Urbani},\ and\ \citenamefont {Zamponi}}]{parisi_theory_2020}%
  \BibitemOpen
  \bibfield  {author} {\bibinfo {author} {\bibfnamefont {G.}~\bibnamefont
  {Parisi}}, \bibinfo {author} {\bibfnamefont {P.}~\bibnamefont {Urbani}}, \
  and\ \bibinfo {author} {\bibfnamefont {F.}~\bibnamefont {Zamponi}},\
  }\href@noop {} {\emph {\bibinfo {title} {Theory of {Simple} {Glasses}:
  {Exact} {Solutions} in {Infinite} {Dimensions}}}}\ (\bibinfo  {publisher}
  {Cambridge University Press},\ \bibinfo {year} {2020})\BibitemShut {NoStop}%
\bibitem [{\citenamefont {Charbonneau}\ \emph {et~al.}(2012)\citenamefont
  {Charbonneau}, \citenamefont {Charbonneau},\ and\ \citenamefont
  {Tarjus}}]{charbonneau_geometrical_2012}%
  \BibitemOpen
  \bibfield  {author} {\bibinfo {author} {\bibfnamefont {B.}~\bibnamefont
  {Charbonneau}}, \bibinfo {author} {\bibfnamefont {P.}~\bibnamefont
  {Charbonneau}}, \ and\ \bibinfo {author} {\bibfnamefont {G.}~\bibnamefont
  {Tarjus}},\ }\href {\doibase 10.1103/PhysRevLett.108.035701} {\bibfield
  {journal} {\bibinfo  {journal} {Phys. Rev. Lett.}\ }\textbf {\bibinfo
  {volume} {108}},\ \bibinfo {pages} {035701} (\bibinfo {year}
  {2012})}\BibitemShut {NoStop}%
\bibitem [{\citenamefont {Charbonneau}\ \emph {et~al.}(2013)\citenamefont
  {Charbonneau}, \citenamefont {Charbonneau},\ and\ \citenamefont
  {Tarjus}}]{charbonneau_geometrical_2013}%
  \BibitemOpen
  \bibfield  {author} {\bibinfo {author} {\bibfnamefont {B.}~\bibnamefont
  {Charbonneau}}, \bibinfo {author} {\bibfnamefont {P.}~\bibnamefont
  {Charbonneau}}, \ and\ \bibinfo {author} {\bibfnamefont {G.}~\bibnamefont
  {Tarjus}},\ }\href {\doibase 10.1063/1.4770498} {\bibfield  {journal}
  {\bibinfo  {journal} {J. Chem. Phys.}\ }\textbf {\bibinfo {volume} {138}},\
  \bibinfo {pages} {12A515} (\bibinfo {year} {2013})}\BibitemShut {NoStop}%
\bibitem [{\citenamefont {Charbonneau}\ \emph {et~al.}(2014)\citenamefont
  {Charbonneau}, \citenamefont {Jin}, \citenamefont {Parisi},\ and\
  \citenamefont {Zamponi}}]{charbonneau_hopping_2014}%
  \BibitemOpen
  \bibfield  {author} {\bibinfo {author} {\bibfnamefont {P.}~\bibnamefont
  {Charbonneau}}, \bibinfo {author} {\bibfnamefont {Y.}~\bibnamefont {Jin}},
  \bibinfo {author} {\bibfnamefont {G.}~\bibnamefont {Parisi}}, \ and\ \bibinfo
  {author} {\bibfnamefont {F.}~\bibnamefont {Zamponi}},\ }\href {\doibase
  10.1073/pnas.1417182111} {\bibfield  {journal} {\bibinfo  {journal} {Proc.
  Natl. Acad. Sci. U.S.A.}\ }\textbf {\bibinfo {volume} {111}},\ \bibinfo
  {pages} {15025} (\bibinfo {year} {2014})}\BibitemShut {NoStop}%
\bibitem [{\citenamefont {Mangeat}\ and\ \citenamefont
  {Zamponi}(2016)}]{mangeat_quantitative_2016}%
  \BibitemOpen
  \bibfield  {author} {\bibinfo {author} {\bibfnamefont {M.}~\bibnamefont
  {Mangeat}}\ and\ \bibinfo {author} {\bibfnamefont {F.}~\bibnamefont
  {Zamponi}},\ }\href {\doibase 10.1103/PhysRevE.93.012609} {\bibfield
  {journal} {\bibinfo  {journal} {Phys. Rev. E}\ }\textbf {\bibinfo {volume}
  {93}},\ \bibinfo {pages} {012609} (\bibinfo {year} {2016})}\BibitemShut
  {NoStop}%
\bibitem [{\citenamefont {van Meel}\ \emph
  {et~al.}(2009{\natexlab{a}})\citenamefont {van Meel}, \citenamefont
  {Frenkel},\ and\ \citenamefont {Charbonneau}}]{van_meel_geometrical_2009}%
  \BibitemOpen
  \bibfield  {author} {\bibinfo {author} {\bibfnamefont {J.~A.}\ \bibnamefont
  {van Meel}}, \bibinfo {author} {\bibfnamefont {D.}~\bibnamefont {Frenkel}}, \
  and\ \bibinfo {author} {\bibfnamefont {P.}~\bibnamefont {Charbonneau}},\
  }\href {\doibase 10.1103/PhysRevE.79.030201} {\bibfield  {journal} {\bibinfo
  {journal} {Phys. Rev. E}\ }\textbf {\bibinfo {volume} {79}},\ \bibinfo
  {pages} {030201} (\bibinfo {year} {2009}{\natexlab{a}})}\BibitemShut
  {NoStop}%
\bibitem [{\citenamefont {van Meel}\ \emph
  {et~al.}(2009{\natexlab{b}})\citenamefont {van Meel}, \citenamefont
  {Charbonneau}, \citenamefont {Fortini},\ and\ \citenamefont
  {Charbonneau}}]{van_meel_hard-sphere_2009}%
  \BibitemOpen
  \bibfield  {author} {\bibinfo {author} {\bibfnamefont {J.~A.}\ \bibnamefont
  {van Meel}}, \bibinfo {author} {\bibfnamefont {B.}~\bibnamefont
  {Charbonneau}}, \bibinfo {author} {\bibfnamefont {A.}~\bibnamefont
  {Fortini}}, \ and\ \bibinfo {author} {\bibfnamefont {P.}~\bibnamefont
  {Charbonneau}},\ }\href {\doibase 10.1103/PhysRevE.80.061110} {\bibfield
  {journal} {\bibinfo  {journal} {Phys. Rev. E}\ }\textbf {\bibinfo {volume}
  {80}},\ \bibinfo {pages} {061110} (\bibinfo {year}
  {2009}{\natexlab{b}})}\BibitemShut {NoStop}%
\bibitem [{\citenamefont {Skoge}\ \emph {et~al.}(2006)\citenamefont {Skoge},
  \citenamefont {Donev}, \citenamefont {Stillinger},\ and\ \citenamefont
  {Torquato}}]{skoge_packing_2006}%
  \BibitemOpen
  \bibfield  {author} {\bibinfo {author} {\bibfnamefont {M.}~\bibnamefont
  {Skoge}}, \bibinfo {author} {\bibfnamefont {A.}~\bibnamefont {Donev}},
  \bibinfo {author} {\bibfnamefont {F.~H.}\ \bibnamefont {Stillinger}}, \ and\
  \bibinfo {author} {\bibfnamefont {S.}~\bibnamefont {Torquato}},\ }\href
  {\doibase 10.1103/PhysRevE.74.041127} {\bibfield  {journal} {\bibinfo
  {journal} {Phys. Rev. E}\ }\textbf {\bibinfo {volume} {74}},\ \bibinfo
  {pages} {041127} (\bibinfo {year} {2006})}\BibitemShut {NoStop}%
\bibitem [{\citenamefont {Charbonneau}\ \emph {et~al.}(2010)\citenamefont
  {Charbonneau}, \citenamefont {Ikeda}, \citenamefont {van Meel},\ and\
  \citenamefont {Miyazaki}}]{charbonneau_numerical_2010}%
  \BibitemOpen
  \bibfield  {author} {\bibinfo {author} {\bibfnamefont {P.}~\bibnamefont
  {Charbonneau}}, \bibinfo {author} {\bibfnamefont {A.}~\bibnamefont {Ikeda}},
  \bibinfo {author} {\bibfnamefont {J.~A.}\ \bibnamefont {van Meel}}, \ and\
  \bibinfo {author} {\bibfnamefont {K.}~\bibnamefont {Miyazaki}},\ }\href
  {\doibase 10.1103/PhysRevE.81.040501} {\bibfield  {journal} {\bibinfo
  {journal} {Phys. Rev. E}\ }\textbf {\bibinfo {volume} {81}},\ \bibinfo
  {pages} {040501} (\bibinfo {year} {2010})}\BibitemShut {NoStop}%
\bibitem [{\citenamefont {Conway}\ and\ \citenamefont
  {Sloane}(1998)}]{conway_sphere_1998}%
  \BibitemOpen
  \bibfield  {author} {\bibinfo {author} {\bibfnamefont {J.}~\bibnamefont
  {Conway}}\ and\ \bibinfo {author} {\bibfnamefont {N.~J.~A.}\ \bibnamefont
  {Sloane}},\ }\href@noop {} {{\selectlanguage {English}\emph {\bibinfo {title}
  {Sphere {Packings}, {Lattices} and {Groups}}}}},\ \bibinfo {edition} {3rd}\
  ed.\ (\bibinfo  {publisher} {Springer},\ \bibinfo {address} {New York},\
  \bibinfo {year} {1998})\BibitemShut {NoStop}%
\bibitem [{\citenamefont {Clisby}\ and\ \citenamefont
  {McCoy}(2006)}]{clisby_ninth_2006}%
  \BibitemOpen
  \bibfield  {author} {\bibinfo {author} {\bibfnamefont {N.}~\bibnamefont
  {Clisby}}\ and\ \bibinfo {author} {\bibfnamefont {B.~M.}\ \bibnamefont
  {McCoy}},\ }\href {\doibase 10.1007/s10955-005-8080-0} {\bibfield  {journal}
  {\bibinfo  {journal} {J. Stat. Phys.}\ }\textbf {\bibinfo {volume} {122}},\
  \bibinfo {pages} {15} (\bibinfo {year} {2006})}\BibitemShut {NoStop}%
\bibitem [{\citenamefont {Bishop}\ and\ \citenamefont
  {Whitlock}(2005)}]{bishop_equation_2005}%
  \BibitemOpen
  \bibfield  {author} {\bibinfo {author} {\bibfnamefont {M.}~\bibnamefont
  {Bishop}}\ and\ \bibinfo {author} {\bibfnamefont {P.~A.}\ \bibnamefont
  {Whitlock}},\ }\href {\doibase 10.1063/1.1874793} {\bibfield  {journal}
  {\bibinfo  {journal} {J. Chem. Phys.}\ }\textbf {\bibinfo {volume} {123}},\
  \bibinfo {pages} {014507} (\bibinfo {year} {2005})}\BibitemShut {NoStop}%
\bibitem [{\citenamefont {Lue}\ and\ \citenamefont
  {Bishop}(2006)}]{lue_molecular_2006}%
  \BibitemOpen
  \bibfield  {author} {\bibinfo {author} {\bibfnamefont {L.}~\bibnamefont
  {Lue}}\ and\ \bibinfo {author} {\bibfnamefont {M.}~\bibnamefont {Bishop}},\
  }\href {\doibase 10.1103/PhysRevE.74.021201} {\bibfield  {journal} {\bibinfo
  {journal} {Phys. Rev. E}\ }\textbf {\bibinfo {volume} {74}},\ \bibinfo
  {pages} {021201} (\bibinfo {year} {2006})}\BibitemShut {NoStop}%
\bibitem [{\citenamefont {Zhang}\ and\ \citenamefont
  {Pettitt}(2014)}]{zhang_computation_2014}%
  \BibitemOpen
  \bibfield  {author} {\bibinfo {author} {\bibfnamefont {C.}~\bibnamefont
  {Zhang}}\ and\ \bibinfo {author} {\bibfnamefont {B.~M.}\ \bibnamefont
  {Pettitt}},\ }\href {\doibase 10.1080/00268976.2014.904945} {\bibfield
  {journal} {\bibinfo  {journal} {Mol. Phys.}\ }\textbf {\bibinfo {volume}
  {112}},\ \bibinfo {pages} {1427} (\bibinfo {year} {2014})}\BibitemShut
  {NoStop}%
\bibitem [{\citenamefont {Frenkel}\ and\ \citenamefont
  {Ladd}(1984)}]{frenkel_new_1984}%
  \BibitemOpen
  \bibfield  {author} {\bibinfo {author} {\bibfnamefont {D.}~\bibnamefont
  {Frenkel}}\ and\ \bibinfo {author} {\bibfnamefont {A.~J.~C.}\ \bibnamefont
  {Ladd}},\ }\href {\doibase 10.1063/1.448024} {\bibfield  {journal} {\bibinfo
  {journal} {J. Chem. Phys.}\ }\textbf {\bibinfo {volume} {81}},\ \bibinfo
  {pages} {3188} (\bibinfo {year} {1984})}\BibitemShut {NoStop}%
\bibitem [{\citenamefont {Frenkel}\ and\ \citenamefont
  {Smit}(2001)}]{frenkel_understanding_2001}%
  \BibitemOpen
  \bibfield  {author} {\bibinfo {author} {\bibfnamefont {D.}~\bibnamefont
  {Frenkel}}\ and\ \bibinfo {author} {\bibfnamefont {B.}~\bibnamefont {Smit}},\
  }\href@noop {} {\emph {\bibinfo {title} {Understanding {Molecular}
  {Simulation}: {From} {Algorithms} to {Applications}}}},\ \bibinfo {edition}
  {2nd}\ ed.\ (\bibinfo  {publisher} {Academic Press},\ \bibinfo {address} {New
  York},\ \bibinfo {year} {2001})\BibitemShut {NoStop}%
\bibitem [{\citenamefont {Khanna}\ \emph {et~al.}(2021)\citenamefont {Khanna},
  \citenamefont {Anwar}, \citenamefont {Frenkel}, \citenamefont {Doherty},\
  and\ \citenamefont {Peters}}]{khanna_free_2021}%
  \BibitemOpen
  \bibfield  {author} {\bibinfo {author} {\bibfnamefont {V.}~\bibnamefont
  {Khanna}}, \bibinfo {author} {\bibfnamefont {J.}~\bibnamefont {Anwar}},
  \bibinfo {author} {\bibfnamefont {D.}~\bibnamefont {Frenkel}}, \bibinfo
  {author} {\bibfnamefont {M.~F.}\ \bibnamefont {Doherty}}, \ and\ \bibinfo
  {author} {\bibfnamefont {B.}~\bibnamefont {Peters}},\ }\href {\doibase
  10.1063/5.0044833} {\bibfield  {journal} {\bibinfo  {journal} {J. Chem.
  Phys.}\ }\textbf {\bibinfo {volume} {154}},\ \bibinfo {pages} {164509}
  (\bibinfo {year} {2021})}\BibitemShut {NoStop}%
\bibitem [{\citenamefont {Best}(1980)}]{best_binary_1980}%
  \BibitemOpen
  \bibfield  {author} {\bibinfo {author} {\bibfnamefont {M.}~\bibnamefont
  {Best}},\ }\href {\doibase 10.1109/TIT.1980.1056269} {\bibfield  {journal}
  {\bibinfo  {journal} {IEEE Trans. Inf. Theory}\ }\textbf {\bibinfo {volume}
  {26}},\ \bibinfo {pages} {738} (\bibinfo {year} {1980})}\BibitemShut
  {NoStop}%
\bibitem [{2dC()}]{2dCrystal}%
  \BibitemOpen
  \href@noop {} {}\bibinfo {note} {We do not here consider the special case
  $d=2$, for which the liquid to solid transition proceeds through a hexatic
  phase, with a weakly first-order liquid-hexatic transition and a continuous
  hexatic-solid transition~\cite{bernard_two-step_2011, engel_hard-disk_2013}.
  Our methodology is indeed not adapted to this situation.}\BibitemShut {Stop}%
\bibitem [{bra()}]{bravais}%
  \BibitemOpen
  \href@noop {} {}\bibinfo {note} {The term Bravais lattice is here used to
  denote what is known in the mathematical literature simply as a lattice.
  Within physics and chemistry communities, $P_{10c}$ could be said to be a
  lattice with a $40$-particle basis, but we instead describe it as a
  non-Bravais-lattice packing to minimize possible confusion with the
  mathematical terminology.}\BibitemShut {Stop}%
\bibitem [{\citenamefont {Coxeter}\ and\ \citenamefont
  {Todd}(1953)}]{coxeter_extreme_1953}%
  \BibitemOpen
  \bibfield  {author} {\bibinfo {author} {\bibfnamefont {H.~S.~M.}\
  \bibnamefont {Coxeter}}\ and\ \bibinfo {author} {\bibfnamefont {J.~A.}\
  \bibnamefont {Todd}},\ }\href {\doibase 10.4153/CJM-1953-043-4} {\bibfield
  {journal} {\bibinfo  {journal} {Canadian Journal of Mathematics}\ }\textbf
  {\bibinfo {volume} {5}},\ \bibinfo {pages} {384} (\bibinfo {year}
  {1953})}\BibitemShut {NoStop}%
\bibitem [{\citenamefont {Conway}\ and\ \citenamefont
  {Sloane}(1982)}]{conway_fast_1982}%
  \BibitemOpen
  \bibfield  {author} {\bibinfo {author} {\bibfnamefont {J.}~\bibnamefont
  {Conway}}\ and\ \bibinfo {author} {\bibfnamefont {N.}~\bibnamefont
  {Sloane}},\ }\href {\doibase 10.1109/TIT.1982.1056484} {\bibfield  {journal}
  {\bibinfo  {journal} {IEEE Trans. Inf. Theory}\ }\textbf {\bibinfo {volume}
  {28}},\ \bibinfo {pages} {227} (\bibinfo {year} {1982})}\BibitemShut
  {NoStop}%
\bibitem [{\citenamefont {Charbonneau}\ \emph
  {et~al.}(2021{\natexlab{a}})\citenamefont {Charbonneau}, \citenamefont {Hu},
  \citenamefont {Kundu},\ and\ \citenamefont
  {Morse}}]{charbonneau_dimensional_2021}%
  \BibitemOpen
  \bibfield  {author} {\bibinfo {author} {\bibfnamefont {P.}~\bibnamefont
  {Charbonneau}}, \bibinfo {author} {\bibfnamefont {Y.}~\bibnamefont {Hu}},
  \bibinfo {author} {\bibfnamefont {J.}~\bibnamefont {Kundu}}, \ and\ \bibinfo
  {author} {\bibfnamefont {P.~K.}\ \bibnamefont {Morse}},\ }\href
  {http://arxiv.org/abs/2111.13749} {\bibfield  {journal} {\bibinfo  {journal}
  {arXiv:2111.13749}\ } (\bibinfo {year} {2021}{\natexlab{a}})}\BibitemShut
  {NoStop}%
\bibitem [{\citenamefont {Charbonneau}\ \emph
  {et~al.}(2021{\natexlab{b}})\citenamefont {Charbonneau}, \citenamefont
  {Morse}, \citenamefont {Perkins},\ and\ \citenamefont
  {Zamponi}}]{charbonneau_three_2021}%
  \BibitemOpen
  \bibfield  {author} {\bibinfo {author} {\bibfnamefont {P.}~\bibnamefont
  {Charbonneau}}, \bibinfo {author} {\bibfnamefont {P.~K.}\ \bibnamefont
  {Morse}}, \bibinfo {author} {\bibfnamefont {W.}~\bibnamefont {Perkins}}, \
  and\ \bibinfo {author} {\bibfnamefont {F.}~\bibnamefont {Zamponi}},\ }\href
  {http://arxiv.org/abs/2109.03063} {\bibfield  {journal} {\bibinfo  {journal}
  {arXiv:2109.03063}\ } (\bibinfo {year} {2021}{\natexlab{b}})}\BibitemShut
  {NoStop}%
\bibitem [{\citenamefont {Pieprzyk}\ \emph {et~al.}(2019)\citenamefont
  {Pieprzyk}, \citenamefont {Bannerman}, \citenamefont {Brańka}, \citenamefont
  {Chudak},\ and\ \citenamefont {Heyes}}]{pieprzyk_thermodynamic_2019}%
  \BibitemOpen
  \bibfield  {author} {\bibinfo {author} {\bibfnamefont {S.}~\bibnamefont
  {Pieprzyk}}, \bibinfo {author} {\bibfnamefont {M.~N.}\ \bibnamefont
  {Bannerman}}, \bibinfo {author} {\bibfnamefont {A.~C.}\ \bibnamefont
  {Brańka}}, \bibinfo {author} {\bibfnamefont {M.}~\bibnamefont {Chudak}}, \
  and\ \bibinfo {author} {\bibfnamefont {D.~M.}\ \bibnamefont {Heyes}},\ }\href
  {\doibase 10.1039/C9CP00903E} {\bibfield  {journal} {\bibinfo  {journal}
  {Phys. Chem. Chem. Phys.}\ }\textbf {\bibinfo {volume} {21}},\ \bibinfo
  {pages} {6886} (\bibinfo {year} {2019})}\BibitemShut {NoStop}%
\bibitem [{\citenamefont {Hall}(1972)}]{hall_another_1972}%
  \BibitemOpen
  \bibfield  {author} {\bibinfo {author} {\bibfnamefont {K.~R.}\ \bibnamefont
  {Hall}},\ }\href {\doibase 10.1063/1.1678576} {\bibfield  {journal} {\bibinfo
   {journal} {J. Chem. Phys.}\ }\textbf {\bibinfo {volume} {57}},\ \bibinfo
  {pages} {2252} (\bibinfo {year} {1972})}\BibitemShut {NoStop}%
\bibitem [{\citenamefont {Speedy}(1998)}]{speedy_pressure_1998}%
  \BibitemOpen
  \bibfield  {author} {\bibinfo {author} {\bibfnamefont {R.~J.}\ \bibnamefont
  {Speedy}},\ }\href {\doibase 10.1088/0953-8984/10/20/006} {\bibfield
  {journal} {\bibinfo  {journal} {J. Phys. Condens. Matter}\ }\textbf {\bibinfo
  {volume} {10}},\ \bibinfo {pages} {4387} (\bibinfo {year}
  {1998})}\BibitemShut {NoStop}%
\bibitem [{\citenamefont {Tarazona}(2000)}]{tarazona_density_2000}%
  \BibitemOpen
  \bibfield  {author} {\bibinfo {author} {\bibfnamefont {P.}~\bibnamefont
  {Tarazona}},\ }\href {\doibase 10.1103/PhysRevLett.84.694} {\bibfield
  {journal} {\bibinfo  {journal} {Phys. Rev. Lett.}\ }\textbf {\bibinfo
  {volume} {84}},\ \bibinfo {pages} {694} (\bibinfo {year} {2000})}\BibitemShut
  {NoStop}%
\bibitem [{\citenamefont {Polson}\ \emph {et~al.}(2000)\citenamefont {Polson},
  \citenamefont {Trizac}, \citenamefont {Pronk},\ and\ \citenamefont
  {Frenkel}}]{polson_finite-size_2000}%
  \BibitemOpen
  \bibfield  {author} {\bibinfo {author} {\bibfnamefont {J.~M.}\ \bibnamefont
  {Polson}}, \bibinfo {author} {\bibfnamefont {E.}~\bibnamefont {Trizac}},
  \bibinfo {author} {\bibfnamefont {S.}~\bibnamefont {Pronk}}, \ and\ \bibinfo
  {author} {\bibfnamefont {D.}~\bibnamefont {Frenkel}},\ }\href {\doibase
  10.1063/1.481102} {\bibfield  {journal} {\bibinfo  {journal} {J. Chem.
  Phys.}\ }\textbf {\bibinfo {volume} {112}},\ \bibinfo {pages} {5339}
  (\bibinfo {year} {2000})}\BibitemShut {NoStop}%
\bibitem [{\citenamefont {Groh}\ and\ \citenamefont
  {Mulder}(2001)}]{groh_closer_2001}%
  \BibitemOpen
  \bibfield  {author} {\bibinfo {author} {\bibfnamefont {B.}~\bibnamefont
  {Groh}}\ and\ \bibinfo {author} {\bibfnamefont {B.}~\bibnamefont {Mulder}},\
  }\href {\doibase 10.1063/1.1342816} {\bibfield  {journal} {\bibinfo
  {journal} {J. Chem. Phys.}\ }\textbf {\bibinfo {volume} {114}},\ \bibinfo
  {pages} {3653} (\bibinfo {year} {2001})}\BibitemShut {NoStop}%
\bibitem [{\citenamefont {Torrie}\ and\ \citenamefont
  {Valleau}(1977)}]{torrie_nonphysical_1977}%
  \BibitemOpen
  \bibfield  {author} {\bibinfo {author} {\bibfnamefont {G.~M.}\ \bibnamefont
  {Torrie}}\ and\ \bibinfo {author} {\bibfnamefont {J.~P.}\ \bibnamefont
  {Valleau}},\ }\href {\doibase 10.1016/0021-9991(77)90121-8} {\bibfield
  {journal} {\bibinfo  {journal} {J. Comput. Phys.}\ }\textbf {\bibinfo
  {volume} {23}},\ \bibinfo {pages} {187} (\bibinfo {year} {1977})}\BibitemShut
  {NoStop}%
\bibitem [{\citenamefont {Charbonneau}\ \emph {et~al.}(2011)\citenamefont
  {Charbonneau}, \citenamefont {Ikeda}, \citenamefont {Parisi},\ and\
  \citenamefont {Zamponi}}]{charbonneau_glass_2011}%
  \BibitemOpen
  \bibfield  {author} {\bibinfo {author} {\bibfnamefont {P.}~\bibnamefont
  {Charbonneau}}, \bibinfo {author} {\bibfnamefont {A.}~\bibnamefont {Ikeda}},
  \bibinfo {author} {\bibfnamefont {G.}~\bibnamefont {Parisi}}, \ and\ \bibinfo
  {author} {\bibfnamefont {F.}~\bibnamefont {Zamponi}},\ }\href {\doibase
  10.1103/PhysRevLett.107.185702} {\bibfield  {journal} {\bibinfo  {journal}
  {Phys. Rev. Lett.}\ }\textbf {\bibinfo {volume} {107}},\ \bibinfo {pages}
  {185702} (\bibinfo {year} {2011})}\BibitemShut {NoStop}%
\bibitem [{\citenamefont {Wilding}\ and\ \citenamefont
  {Bruce}(2000)}]{wilding_freezing_2000}%
  \BibitemOpen
  \bibfield  {author} {\bibinfo {author} {\bibfnamefont {N.~B.}\ \bibnamefont
  {Wilding}}\ and\ \bibinfo {author} {\bibfnamefont {A.~D.}\ \bibnamefont
  {Bruce}},\ }\href {\doibase 10.1103/PhysRevLett.85.5138} {\bibfield
  {journal} {\bibinfo  {journal} {Phys. Rev. Lett.}\ }\textbf {\bibinfo
  {volume} {85}},\ \bibinfo {pages} {5138} (\bibinfo {year}
  {2000})}\BibitemShut {NoStop}%
\bibitem [{\citenamefont {Wilding}(2002)}]{wilding_new_2002}%
  \BibitemOpen
  \bibfield  {author} {\bibinfo {author} {\bibfnamefont {N.~B.}\ \bibnamefont
  {Wilding}},\ }\href {\doibase 10.1016/S0010-4655(02)00440-X} {\bibfield
  {journal} {\bibinfo  {journal} {Comput. Phys. Commun.}\ }\textbf {\bibinfo
  {volume} {146}},\ \bibinfo {pages} {99} (\bibinfo {year} {2002})}\BibitemShut
  {NoStop}%
\bibitem [{\citenamefont {Engel}\ \emph {et~al.}(2013)\citenamefont {Engel},
  \citenamefont {Anderson}, \citenamefont {Glotzer}, \citenamefont {Isobe},
  \citenamefont {Bernard},\ and\ \citenamefont
  {Krauth}}]{engel_hard-disk_2013}%
  \BibitemOpen
  \bibfield  {author} {\bibinfo {author} {\bibfnamefont {M.}~\bibnamefont
  {Engel}}, \bibinfo {author} {\bibfnamefont {J.~A.}\ \bibnamefont {Anderson}},
  \bibinfo {author} {\bibfnamefont {S.~C.}\ \bibnamefont {Glotzer}}, \bibinfo
  {author} {\bibfnamefont {M.}~\bibnamefont {Isobe}}, \bibinfo {author}
  {\bibfnamefont {E.~P.}\ \bibnamefont {Bernard}}, \ and\ \bibinfo {author}
  {\bibfnamefont {W.}~\bibnamefont {Krauth}},\ }\href {\doibase
  10.1103/PhysRevE.87.042134} {\bibfield  {journal} {\bibinfo  {journal} {Phys.
  Rev. E}\ }\textbf {\bibinfo {volume} {87}},\ \bibinfo {pages} {042134}
  (\bibinfo {year} {2013})}\BibitemShut {NoStop}%
\bibitem [{\citenamefont {Ruiz-Franco}\ \emph {et~al.}(2019)\citenamefont
  {Ruiz-Franco}, \citenamefont {Zaccarelli}, \citenamefont {Schöpe},\ and\
  \citenamefont {van Megen}}]{ruiz-franco_coincidence_2019}%
  \BibitemOpen
  \bibfield  {author} {\bibinfo {author} {\bibfnamefont {J.}~\bibnamefont
  {Ruiz-Franco}}, \bibinfo {author} {\bibfnamefont {E.}~\bibnamefont
  {Zaccarelli}}, \bibinfo {author} {\bibfnamefont {H.~J.}\ \bibnamefont
  {Schöpe}}, \ and\ \bibinfo {author} {\bibfnamefont {W.}~\bibnamefont {van
  Megen}},\ }\href {\doibase 10.1063/1.5114720} {\bibfield  {journal} {\bibinfo
   {journal} {J. Chem. Phys.}\ }\textbf {\bibinfo {volume} {151}},\ \bibinfo
  {pages} {104501} (\bibinfo {year} {2019})}\BibitemShut {NoStop}%
\bibitem [{\citenamefont {Wang}(2005)}]{wang_mean-field_2005}%
  \BibitemOpen
  \bibfield  {author} {\bibinfo {author} {\bibfnamefont {X.-Z.}\ \bibnamefont
  {Wang}},\ }\href {\doibase 10.1063/1.1840444} {\bibfield  {journal} {\bibinfo
   {journal} {J. Chem. Phys.}\ }\textbf {\bibinfo {volume} {122}},\ \bibinfo
  {pages} {044515} (\bibinfo {year} {2005})}\BibitemShut {NoStop}%
\bibitem [{\citenamefont {Rogers}(1958)}]{rogers_packing_1958}%
  \BibitemOpen
  \bibfield  {author} {\bibinfo {author} {\bibfnamefont {C.~A.}\ \bibnamefont
  {Rogers}},\ }\href {\doibase https://doi.org/10.1112/plms/s3-8.4.609}
  {\bibfield  {journal} {\bibinfo  {journal} {Proc. London Math. Soc.}\
  }\textbf {\bibinfo {volume} {s3-8}},\ \bibinfo {pages} {609} (\bibinfo {year}
  {1958})}\BibitemShut {NoStop}%
\bibitem [{\citenamefont {Cohn}\ and\ \citenamefont
  {Elkies}(2003)}]{cohn_new_2003}%
  \BibitemOpen
  \bibfield  {author} {\bibinfo {author} {\bibfnamefont {H.}~\bibnamefont
  {Cohn}}\ and\ \bibinfo {author} {\bibfnamefont {N.}~\bibnamefont {Elkies}},\
  }\href {\doibase 10.4007/annals.2003.157.689} {\bibfield  {journal} {\bibinfo
   {journal} {Ann. Math.}\ }\textbf {\bibinfo {volume} {157}},\ \bibinfo
  {pages} {689} (\bibinfo {year} {2003})}\BibitemShut {NoStop}%
\bibitem [{\citenamefont {Viazovska}(2017)}]{viazovska_sphere_2017}%
  \BibitemOpen
  \bibfield  {author} {\bibinfo {author} {\bibfnamefont {M.~S.}\ \bibnamefont
  {Viazovska}},\ }\href {https://www.jstor.org/stable/26395747} {\bibfield
  {journal} {\bibinfo  {journal} {Ann. Math.}\ }\textbf {\bibinfo {volume}
  {185}},\ \bibinfo {pages} {991} (\bibinfo {year} {2017})}\BibitemShut
  {NoStop}%
\bibitem [{\citenamefont {Cohn}(2017)}]{cohn_conceptual_2017}%
  \BibitemOpen
  \bibfield  {author} {\bibinfo {author} {\bibfnamefont {H.}~\bibnamefont
  {Cohn}},\ }\href {\doibase 10.1090/noti1474} {\bibfield  {journal} {\bibinfo
  {journal} {Not. Am. Math. Soc.}\ }\textbf {\bibinfo {volume} {64}},\ \bibinfo
  {pages} {102} (\bibinfo {year} {2017})}\BibitemShut {NoStop}%
\bibitem [{\citenamefont {Estrada}\ and\ \citenamefont
  {Robles}(2011)}]{estrada_fluidsolid_2011}%
  \BibitemOpen
  \bibfield  {author} {\bibinfo {author} {\bibfnamefont {C.~D.}\ \bibnamefont
  {Estrada}}\ and\ \bibinfo {author} {\bibfnamefont {M.}~\bibnamefont
  {Robles}},\ }\href {\doibase 10.1063/1.3530780} {\bibfield  {journal}
  {\bibinfo  {journal} {J. Chem. Phys.}\ }\textbf {\bibinfo {volume} {134}},\
  \bibinfo {pages} {044115} (\bibinfo {year} {2011})}\BibitemShut {NoStop}%
\bibitem [{\citenamefont {Hoover}\ and\ \citenamefont
  {Ree}(1968)}]{hoover_melting_1968}%
  \BibitemOpen
  \bibfield  {author} {\bibinfo {author} {\bibfnamefont {W.~G.}\ \bibnamefont
  {Hoover}}\ and\ \bibinfo {author} {\bibfnamefont {F.~H.}\ \bibnamefont
  {Ree}},\ }\href {\doibase 10.1063/1.1670641} {\bibfield  {journal} {\bibinfo
  {journal} {J. Chem. Phys.}\ }\textbf {\bibinfo {volume} {49}},\ \bibinfo
  {pages} {3609} (\bibinfo {year} {1968})}\BibitemShut {NoStop}%
\bibitem [{\citenamefont {Michels}\ and\ \citenamefont
  {Trappeniers}(1984)}]{michels_dynamical_1984}%
  \BibitemOpen
  \bibfield  {author} {\bibinfo {author} {\bibfnamefont {J.~P.~J.}\
  \bibnamefont {Michels}}\ and\ \bibinfo {author} {\bibfnamefont {N.~J.}\
  \bibnamefont {Trappeniers}},\ }\href {\doibase 10.1016/0375-9601(84)90749-7}
  {\bibfield  {journal} {\bibinfo  {journal} {Phys. Lett. A}\ }\textbf
  {\bibinfo {volume} {104}},\ \bibinfo {pages} {425} (\bibinfo {year}
  {1984})}\BibitemShut {NoStop}%
\bibitem [{\citenamefont {Speedy}(1997)}]{speedy_pressure_1997}%
  \BibitemOpen
  \bibfield  {author} {\bibinfo {author} {\bibfnamefont {R.~J.}\ \bibnamefont
  {Speedy}},\ }\href {\doibase 10.1088/0953-8984/9/41/006} {\bibfield
  {journal} {\bibinfo  {journal} {J. Phys.: Condens. Matter}\ }\textbf
  {\bibinfo {volume} {9}},\ \bibinfo {pages} {8591} (\bibinfo {year}
  {1997})}\BibitemShut {NoStop}%
\bibitem [{\citenamefont {Alder}\ and\ \citenamefont
  {Wainwright}(1957)}]{alder_phase_1957}%
  \BibitemOpen
  \bibfield  {author} {\bibinfo {author} {\bibfnamefont {B.~J.}\ \bibnamefont
  {Alder}}\ and\ \bibinfo {author} {\bibfnamefont {T.~E.}\ \bibnamefont
  {Wainwright}},\ }\href {\doibase 10.1063/1.1743957} {\bibfield  {journal}
  {\bibinfo  {journal} {J. Chem. Phys.}\ }\textbf {\bibinfo {volume} {27}},\
  \bibinfo {pages} {1208} (\bibinfo {year} {1957})}\BibitemShut {NoStop}%
\bibitem [{\citenamefont {Bryant}\ \emph {et~al.}(2002)\citenamefont {Bryant},
  \citenamefont {Williams}, \citenamefont {Qian}, \citenamefont {Snook},
  \citenamefont {Perez},\ and\ \citenamefont {Pincet}}]{bryant_how_2002}%
  \BibitemOpen
  \bibfield  {author} {\bibinfo {author} {\bibfnamefont {G.}~\bibnamefont
  {Bryant}}, \bibinfo {author} {\bibfnamefont {S.~R.}\ \bibnamefont
  {Williams}}, \bibinfo {author} {\bibfnamefont {L.}~\bibnamefont {Qian}},
  \bibinfo {author} {\bibfnamefont {I.~K.}\ \bibnamefont {Snook}}, \bibinfo
  {author} {\bibfnamefont {E.}~\bibnamefont {Perez}}, \ and\ \bibinfo {author}
  {\bibfnamefont {F.}~\bibnamefont {Pincet}},\ }\href {\doibase
  10.1103/PhysRevE.66.060501} {\bibfield  {journal} {\bibinfo  {journal} {Phys.
  Rev. E}\ }\textbf {\bibinfo {volume} {66}},\ \bibinfo {pages} {060501}
  (\bibinfo {year} {2002})}\BibitemShut {NoStop}%
\bibitem [{\citenamefont {Vega}\ and\ \citenamefont
  {Noya}(2007)}]{vega_revisiting_2007}%
  \BibitemOpen
  \bibfield  {author} {\bibinfo {author} {\bibfnamefont {C.}~\bibnamefont
  {Vega}}\ and\ \bibinfo {author} {\bibfnamefont {E.~G.}\ \bibnamefont
  {Noya}},\ }\href {\doibase 10.1063/1.2790426} {\bibfield  {journal} {\bibinfo
   {journal} {J. Chem. Phys.}\ }\textbf {\bibinfo {volume} {127}},\ \bibinfo
  {pages} {154113} (\bibinfo {year} {2007})}\BibitemShut {NoStop}%
\bibitem [{\citenamefont {Richard}\ and\ \citenamefont
  {Speck}(2018{\natexlab{a}})}]{richard_crystallization_2018}%
  \BibitemOpen
  \bibfield  {author} {\bibinfo {author} {\bibfnamefont {D.}~\bibnamefont
  {Richard}}\ and\ \bibinfo {author} {\bibfnamefont {T.}~\bibnamefont
  {Speck}},\ }\href {\doibase 10.1063/1.5025394} {\bibfield  {journal}
  {\bibinfo  {journal} {J. Chem. Phys.}\ }\textbf {\bibinfo {volume} {148}},\
  \bibinfo {pages} {224102} (\bibinfo {year} {2018}{\natexlab{a}})}\BibitemShut
  {NoStop}%
\bibitem [{\citenamefont {Richard}\ and\ \citenamefont
  {Speck}(2018{\natexlab{b}})}]{richard_crystallization_2018-1}%
  \BibitemOpen
  \bibfield  {author} {\bibinfo {author} {\bibfnamefont {D.}~\bibnamefont
  {Richard}}\ and\ \bibinfo {author} {\bibfnamefont {T.}~\bibnamefont
  {Speck}},\ }\href {\doibase 10.1063/1.5016277} {\bibfield  {journal}
  {\bibinfo  {journal} {J. Chem. Phys.}\ }\textbf {\bibinfo {volume} {148}},\
  \bibinfo {pages} {124110} (\bibinfo {year} {2018}{\natexlab{b}})}\BibitemShut
  {NoStop}%
\bibitem [{\citenamefont {Bültmann}\ and\ \citenamefont
  {Schilling}(2020)}]{Bultmann_computation_2020}%
  \BibitemOpen
  \bibfield  {author} {\bibinfo {author} {\bibfnamefont {M.}~\bibnamefont
  {Bültmann}}\ and\ \bibinfo {author} {\bibfnamefont {T.}~\bibnamefont
  {Schilling}},\ }\href {\doibase 10.1103/PhysRevE.102.042123} {\bibfield
  {journal} {\bibinfo  {journal} {Phys. Rev. E}\ }\textbf {\bibinfo {volume}
  {102}},\ \bibinfo {pages} {042123} (\bibinfo {year} {2020})}\BibitemShut
  {NoStop}%
\bibitem [{\citenamefont {Damasceno}\ \emph {et~al.}(2012)\citenamefont
  {Damasceno}, \citenamefont {Engel},\ and\ \citenamefont
  {Glotzer}}]{damasceno_predictive_2012}%
  \BibitemOpen
  \bibfield  {author} {\bibinfo {author} {\bibfnamefont {P.~F.}\ \bibnamefont
  {Damasceno}}, \bibinfo {author} {\bibfnamefont {M.}~\bibnamefont {Engel}}, \
  and\ \bibinfo {author} {\bibfnamefont {S.~C.}\ \bibnamefont {Glotzer}},\
  }\href {\doibase 10.1126/science.1220869} {\bibfield  {journal} {\bibinfo
  {journal} {Science}\ }\textbf {\bibinfo {volume} {337}},\ \bibinfo {pages}
  {453} (\bibinfo {year} {2012})}\BibitemShut {NoStop}%
\bibitem [{\citenamefont {Batten}\ \emph {et~al.}(2010)\citenamefont {Batten},
  \citenamefont {Stillinger},\ and\ \citenamefont
  {Torquato}}]{batten_phase_2010}%
  \BibitemOpen
  \bibfield  {author} {\bibinfo {author} {\bibfnamefont {R.~D.}\ \bibnamefont
  {Batten}}, \bibinfo {author} {\bibfnamefont {F.~H.}\ \bibnamefont
  {Stillinger}}, \ and\ \bibinfo {author} {\bibfnamefont {S.}~\bibnamefont
  {Torquato}},\ }\href {\doibase 10.1103/PhysRevE.81.061105} {\bibfield
  {journal} {\bibinfo  {journal} {Phys. Rev. E}\ }\textbf {\bibinfo {volume}
  {81}},\ \bibinfo {pages} {061105} (\bibinfo {year} {2010})}\BibitemShut
  {NoStop}%
\bibitem [{dat()}]{data}%
  \BibitemOpen
  \href@noop {} {}\bibinfo {note} {Duke digital repository,
  \href{https://doi.org/10.7924/r4jh3mw3w}{DOI: 10.7924/r4jh3mw3w}}\BibitemShut
  {NoStop}%
\bibitem [{\citenamefont {Bernard}\ and\ \citenamefont
  {Krauth}(2011)}]{bernard_two-step_2011}%
  \BibitemOpen
  \bibfield  {author} {\bibinfo {author} {\bibfnamefont {E.~P.}\ \bibnamefont
  {Bernard}}\ and\ \bibinfo {author} {\bibfnamefont {W.}~\bibnamefont
  {Krauth}},\ }\href {\doibase 10.1103/PhysRevLett.107.155704} {\bibfield
  {journal} {\bibinfo  {journal} {Phys. Rev. Lett.}\ }\textbf {\bibinfo
  {volume} {107}},\ \bibinfo {pages} {155704} (\bibinfo {year}
  {2011})}\BibitemShut {NoStop}%
\end{thebibliography}%


%

\clearpage

\end{document}